\documentclass[aip,jcp,reprint,showkeys,groupedaddress,floatfix]{revtex4-1}
\pdfoutput=1

\usepackage[sort&compress]{natbib}
\usepackage{graphicx,color}
\usepackage{rotating}
\usepackage{physics,amssymb}

\newcommand{\molpro}{\texttt{MOLPRO}}
\newcommand{\duo}{\texttt{DUO}}
\newcommand{\wn}{~cm\(^{-1}\)}
\newcommand{\Xst}{\text{X} \ensuremath{^2\Sigma^+}}
\newcommand{\Hst}{\text{H} \ensuremath{^2\Delta}}
\newcommand{\Ast}{\text{A} \ensuremath{^2\Pi}}
\newcommand{\Bst}{\text{B} \ensuremath{^2\Sigma^+}}
\newcommand{\Est}{\text{E} \ensuremath{^2\Pi}}

\newcommand{\Dst}{\text{D} \ensuremath{^2\Sigma^+}}

\begin{document}

\title{Quantitative theoretical analysis of lifetimes and decay rates relevant in laser cooling BaH}

\newcommand{\qub}{
	School of Chemistry and Chemical Engineering,
	Queen's University Belfast,
	Stranmillis Road,
	Belfast BT9 5AG,
	UK}
%\affiliation{\qub}
\author{Keith Moore}
%\affiliation{\qub}
\author{Ian C. Lane}
\email[Corresponding author: ]{i.lane@qub.ac.uk}
\affiliation{\qub}

%\corref{cor1}}
%\email{i.lane@qub.ac.uk} \cortext[cor1] {Corresponding author.}

\begin{abstract}
Tiny radiative losses below the 0.1\% level can prove ruinous to the effective laser cooling of a molecule.
In this paper the laser cooling of a hydride is studied with rovibronic detail using {\it ab initio} quantum chemistry in order to document the decays to all possible electronic states (not just the vibrational branching within a single electronic transition) and to identify the most populated final quantum states.
The effect of spin-orbit and associated couplings on the properties of the lowest excited states of BaH are analysed in detail.
The lifetimes of the A$^2\Pi_{1/2}$, H$^2\Delta_{3/2}$ and E$^2\Pi_{1/2}$ states are calculated (136~ns, 5.8~$\mu$s and 46~ns respectively) for the first time, while the theoretical value for B$^2\Sigma^{+}_{1/2}$ is in good agreement with experiments.
Using a simple rate model the numbers of absorption-emission cycles possible for both one- and two-colour cooling on the competing electronic transitions are determined, and it is clearly demonstrated that the A$^2\Pi$ -- X$^2\Sigma^+$ transition is superior to B$^2\Sigma^+$ -- X$^2\Sigma^+$, where multiple tiny decay channels degrade its efficiency.
Further possible improvements to the cooling method are proposed.
\end{abstract}


\maketitle

%----------------------------------------------------------------
\section{
\label{sec:intro}
Introduction}
%----------------------------------------------------------------

The BaH molecule is an intriguing \cite{Lane2015} laser cooling candidate, partly because it would be the first hydride to be cooled in this way but also because it is a potential source of ultracold hydrogen atoms.
Along with its sister molecules BeH \cite{Dattani2015} and MgH \cite{Henderson2013}, this radical is one of the most studied metal-bearing diatomic hydrides since it's first laboratory identification in 1909~\cite{Eagle1909}.
It possesses three low lying excited states correlated to the 5$d$ state of the Ba atom that share very similar spectroscopic parameters that closely resemble the X$^2\Sigma^+$ ground state.
This ensures diagonal Franck-Condon (FC) factors close to 1 that are crucial to ensuring the enormous number of absorption-emission cycles (over 10$^4$) required for strong laser cooling with a manageable number of light fields.
Indeed, because of the large vibrational separation in hydrides BaH appears to be one of the few molecular systems where greater than 5~000 cycles \cite{Lane2015} can be achieved with just two vibronic transitions (lasers).
In addition, all three 5$d$ states and even the higher lying E$^2\Pi$ curve lie below the first dissociation limit, ensuring that predissociation is not a loss mechanism.
This is not the case with the lighter alkaline-earth hydrides, weakening the ability of laser methods \cite{Gao2014} to effectively cool those systems.
Already a buffer gas beam \cite{Hutzler2012} of BaH molecules in the  X$^2\Sigma^+_{1/2}$ ($v^{\prime\prime}$~=~0, $N^{\prime\prime}$~=~1) state has been tested \cite{Tarallo2016, Iwata2017} in preparation for future laser cooling experiments.

The optical and near-infrared spectra \cite{Watson1933, Watson1935, Kopp1966-1, Kopp1966-2, Appelblad1985, Fabre1987, Bernard1987, Bernard1989, Barrow1991, Walker1993, Berg1997, Ram2013} of BaH is dominated by the three 5$d$-complex states.
The lowest excited electronic H$^2\Delta$ state was first discovered as a perturbating state in the near infra-red spectrum \cite{Watson1935} of BaH.
It was finally directly measured \cite{Fabre1987} in an laser-based study in 1987, though only the upper spin-orbit component was reported.
It wasn't long, however, before the H$^2\Delta_{3/2}$ state was observed when 52 lines were recorded in a thermal (1000 $^\circ$C) emission study subsequently included in a global fit \cite{Bernard1987} of the BaH visible and near infra-red spectra.
These lines belonged to the H$^2\Delta_{3/2}$ -- X$^2\Sigma^+_{1/2}$ (0-0) band system.
The equivalent spectrum, however, in BaD was much less intense, suggesting a very weak transition.
A later, more comprehensive analysis \cite{Bernard1989} included Laser Induced Fluorescence (LIF) transitions involving H$^2\Delta_{3/2}$ $v$ = 1.
This work folded in spectroscopic data involving all three 5$d$-complex states in order to derive the spectroscopic constants and is currently the only experimental measurement of the spin-orbit splitting ($A$ = 217.298~cm$^{-1}$ for $\nu^{\prime}$ = 0) in the H$^2\Delta$ state.
A final study \cite{Barrow1991} of the H$^2\Delta$ -- X$^2\Sigma^+$ chemiluminescence, including observations at higher temperatures, attempted to explain some anomalies in the observed spectrum.

H$^2\Delta$ perturbs the strong A$^2\Pi $ -- X$^2\Sigma^+$ spectrum first studied by Watson during the 1930s \cite{Watson1933} and then thirty years later by the group led by Kopp \cite{Kopp1966-2}.
 There is currently a significant discrepancy in experimental constants for the A$^2\Pi$ spin-orbit separation, with Kopp {\it et al.} \cite{Kopp1966-2} determining this to be around 483~cm$^{-1}$ for $\nu^{\prime}$ = 0 while Barrow and co-workers quote a fitted fine-structure constant \cite{Bernard1989} of $A$~=~341.2~cm$^{-1}$ for the same vibrational level.
Furthermore, neither the lifetime of the A$^2\Pi$ nor H$^2\Delta$ state have been measured in experiments.

The B$^2\Sigma^+ $ -- X$^2\Sigma^+$ bands of BaH and BaD were reported \cite{Watson1933} again by Watson while similar work \cite{Kopp1966-2} on BaD was published by Kopp {\it et al.} some thirty years later.
Later high-quality Fourier-transform data \cite{Appelblad1985} of BaH provided some improvement
in wavenumber measurements and spectroscopic constants for the B$^2\Sigma^+$ $v^{\prime}$= 0-3 vibronic levels.
Members of that experimental team later determined the lifetime \cite{Berg1997} of the B$^2\Sigma^+$ ($v^{\prime}$ = 0, $J$ = 11/2) level as 124~$\pm$~2~ns.
Such a measurement is of huge importance to the ultracold molecule community because the rate of laser cooling is crucial in maximising the efficiency of the process.
Too long an excited state lifetime and the cooling process is too slow, but a very short lifetime results in a high Doppler temperature $T_D$ (this is determined by the natural linewidth).
A further concern is the unwelcome presence of additional radiative decay channels that will ensure the laser cooling cycle ultimately terminates:
any lower lying state that can optically couple with the upper cooling level can potentially induce significant losses.

The highly diagonal nature of the lowest energy electronic transitions has somewhat restricted the information that experiments can reveal on the bond length dependence of the electronic energies.
Fortunately, theoretical quantum chemistry can help fill these gaps and provide information of a number of key properties such as the dissociation energy.
A recently calculated potential \cite{Moore2016} that incorporated both {\it ab initio} and experimental data matches the lowest vibrational levels with sub-wavenumber accuracy.
However, there remains only a single theoretical \cite{Allouche1992} study that includes the effects of spin-orbit interactions, though the obtained spin-orbit coupling (SO-coupling) constants for the $^2\Pi$ states were rather higher than the experimental values.
{\it Ab initio} calculations have also been shown \cite{Wells2011} to be a useful method to screen for suitable laser cooling candidates.
Previous work \cite{Lane2015,Gao2014} on laser cooling in BaH has not, however, included the effect of the unpaired electron spin on the electronic states and ro-vibrational levels.
BaH is a particularly attractive diatomic for a more detailed, quantitative description of the laser cooling process because a number of the important physical parameters have also been measured (summarised in Table~\ref{tab:bah_exp_data}) that can help refine the theoretical calculations.
However, key information is still missing from the data that theoretical work can help provide. 
The present contribution aims to produce the most complete theoretical analysis to date on laser cooling a diatomic, and in particular of the BaH radical and the associated electronic transitions.

%Useful experimental data
\begin{table}
\caption{\label{tab:bah_exp_data}Experimental data on BaH useful for laser cooling studies.}
\small\setlength\tabcolsep{5pt}\center\begin{tabular}{lll}

\hline
Measured property & states    & Refs.\\\hline
\noalign{\vskip 1.1mm}
$r_e$
  & X, H\textsuperscript{a}, A   & \cite{Ram2013}, \cite{Bernard1989}, \cite{Kopp1966-1} \\
  & B, E    & \cite{Appelblad1985}, \cite{Ram2013}\\
$T_{00}$ and $T_{10}$
  & H\textsuperscript{b}, A   & \cite{Fabre1987}, \cite{Kopp1966-1} \\
  & B, E    & \cite{Appelblad1985}, \cite{Ram2013}\\  
$v$ = 2
  & X, A    & \cite{Ram2013}, \cite{Kopp1966-1} \\
  & B, E    & \cite{Appelblad1985}, \cite{Ram2013}\\      
$v$ = 3 
  & X, B   & \cite{Walker1993}, \cite{Appelblad1985} \\
  & E      & \cite{Ram2013} \\   
Spin-orbit separation
  & H, A\textsuperscript{c} & \cite{Bernard1989}, \cite{Kopp1966-1} \\
  & E      & \cite{Ram2013} \\ 
Lifetime
  & B   & \cite{Berg1997}   \\ 
\hline
\noalign{\vskip 1.5mm}
\multicolumn{3}{l}{\textsuperscript{a} Deperturbed value \cite{Bernard1989} quoted.}\\
\multicolumn{3}{l}{\textsuperscript{b} Only H$^2\Delta_{5/2}$ reported in \cite{Fabre1987}. }\\
\multicolumn{3}{l}{\textsuperscript{c} The A$^2\Pi$ state separation of \cite{Bernard1989} and  \cite{Kopp1966-1} are different.}\\
\end{tabular}
\end{table}

%----------------------------------------------------------------
\section{
\label{sec:ai_calc}
\emph{Ab initio} calculations}
%----------------------------------------------------------------

{\it Ab initio} calculations of the potential energy curves were performed at a post Hartree-Fock level using a parallel version of the 
 MOLPRO \cite{Werner2010} (version 2010.1) suite of quantum chemistry codes.
An earlier study had demonstrated that describing the barium atomic orbitals using the aug-cc-pCV$n$Z basis sets ($n$ = $Q$,5) taken to the CBS (Complete Basis Set) limit could produce \cite{Moore2016} a highly accurate ground state potential for BaH.
In the present work the smaller aug-cc-pCV$Q$Z (ACV$Q$Z) basis set \cite{Li2013} was used on the barium atom (to describe the 5$s$5$p$6$s$ electrons) and the study extended to include the excited molecular electronic states. An effective core potential (ECP) is used \cite{Lim2006} to describe the lowest 46-core electrons. The ACV$Q$Z basis set was used as it is a good compromise of accuracy and computational speed \cite{Moore2016} and was produced by taking the most diffuse exponents from the AV$Q$Z barium basis set and adding them to the standard \cite{Li2013} cc-pCV$Q$Z functions.
To describe the atomic orbitals on the hydrogen atom the equivalent aug-cc-pV$Q$Z basis was used~\cite{Dunning1989}.
The active space consisted of all the occupied valence orbitals plus the 6$p$5$d$ and the lowest Rydberg 7$s$-orbital on barium (three electrons in eleven orbitals). The electron correlation was determined using both the State-Averaged Complete Active Space Self-Consistent Field~\cite{Siegbahn1980} (SA-CASSCF) and the Multi-reference Configuration Interaction~\cite{Knowles1988} (MRCI) methods (for static and dynamic correlation, respectively).
The latter is restricted to excitations of single or double electrons (three electrons in eleven orbitals) so to estimate the higher order contributions the Davidson correction~\cite{Davidson1974} was applied. The Abelian point group C$_{2v}$ is used to describe the diatomic orbitals and symmetry labels.
 All doublet states expected from the first five atomic limits were included in the SA-CASSCF calculation, namely 5\(\times ^2\Sigma^+\), 4\(\times ^2\Pi\), 2\(\times ^2\Delta \equiv \) 7\(\times ^2A_1\), 4\(\times ^2B_1\), 4\(\times ^2B_2\), 2\(\times ^2A_2\) (abbreviated to CAS-7442), while only the 5 states considered in this study were included in the subsequent MRCI calculations (\Xst, \Hst, \Ast, \Bst, \Est\ \(\equiv \) 3\(\times ^2A_1\), 2\(\times ^2B_1\), 2\(\times ^2B_2\), 1\(\times ^2A_2 \equiv \) MRCI-3221).
The resulting potentials are in good agreement with previous work from this group using the aug-cc-pV$6$Z basis~\cite{Moore2016} set.

Next, the SO-couplings present were determined~\cite{Berning2000} using the same basis set and again the MOLPRO suite of programs.
In this paper we adopt the traditional Hund's (a) electronic label for those potentials calculated without consideration of spin-orbit coupling effects such as A$^2\Pi$ whilst using the form
B$^2\Sigma^{+}_{1/2}$ for the final states where $\Omega$ (the projection of the total electronic angular momentum on the internuclear axis) is a good quantum number.
This is consistent with the observation \cite{Allouche1992} that even after the inclusion of mixing, the states considered here retain their fundamental symmetry character in the Franck-Condon region.

% MRCIQ Equilibrium Values
\begin{table}
\center
\caption{Equilibrium distance and energy values for electronic states of BaH, as determined by spline interpolation of the MRCI+Q {\it ab initio} points (ACV$Q$Z basis set). $\Delta$ is the difference between the calculated and experimental values.}
\label{tab:bah_ciq_eqval}
\center
\footnotesize
\begin{tabular}{|l|rr|rr|}
\noalign{\vskip 2.6mm}
\hline
State	& \(r_e\)/\AA & \(\Delta_{r_e}\)/\AA
		& \(T_e\)/cm\(^{-1}\) & \(\Delta_{T_e}\)/cm\(^{-1}\) \rule{0pt}{2.5ex}
		%& \(D_e\)/cm\(^{-1}\) 
\\[.6ex]\hline
\Xst
	& 2.239 & +0.007 
	& 0.00 & --
%	& 16708
	\rule{0pt}{2.5ex}\\
	&&  (+0.3\%)  &&\\[1ex]
\Hst
	& 2.295 & +0.007 
	& 9698.98 & +491.37 
%	& 16509
	\rule{0pt}{2.5ex}\\
	&&  (+0.3\%)  &&  (+5.3\%)  \\[1ex]
\Ast
	& 2.279 & +0.019 
	& 10076.45 & +377.82 
%	& 16132
	\rule{0pt}{2.5ex}\\
	&&  (+0.9\%)  &&  (+3.9\%)  \\[1ex]
\Bst
	& 2.291 & +0.023
	& 11112.61 & +20.02
%	& 16996
	\rule{0pt}{2.5ex}\\
	&&  (+1.0\%)  && (+0.2\%)  \\[1ex]
\Est
	& 2.190 & +0.002 
	& 14871.07 & +40.91 
%	& 13237
	\rule{0pt}{2.5ex}\\
	&& (+0.1\%)  &&  (+0.3\%)  \\[0.8ex]\hline
\end{tabular}
\end{table}

%----------------------------------------------------------------
\section{
\label{sec:theorres}
Theoretical results}
%----------------------------------------------------------------

\subsection{Potential energy curves}\label{PECurves}
Particular care was required in converting the \(A_1\) symmetry states to the true electronic states, since root flipping in the symmetry repeatedly leads to the identity of the state switching between \Hst, \Bst\ and \Dst.
As such, sections of the \Hst\ are determined using 2\(A_1\) \&  1\(A_2\), 3\(A_1\) \& 1\(A_2\) or solely 1\(A_2\).
Similarly, a section of \Bst\ is undetermined (because at extended bond lengths 3\(A_1\) is actually \Dst ) but this is well outside the FC region (a spline interpolation is used in the figures).
Resolution of these states may be better handled in the future through inclusion of a 4\textsuperscript{th} \(A_1\) state but the aim here was to describe the FC region using the fewest possible states in the MRCI calculation.
By contrast the process of combining \(B_1\) and \(B_2\) representations of \(^2\Pi\) states is very straightforward.

\begin{figure*}
\centering
\includegraphics[width=140mm]{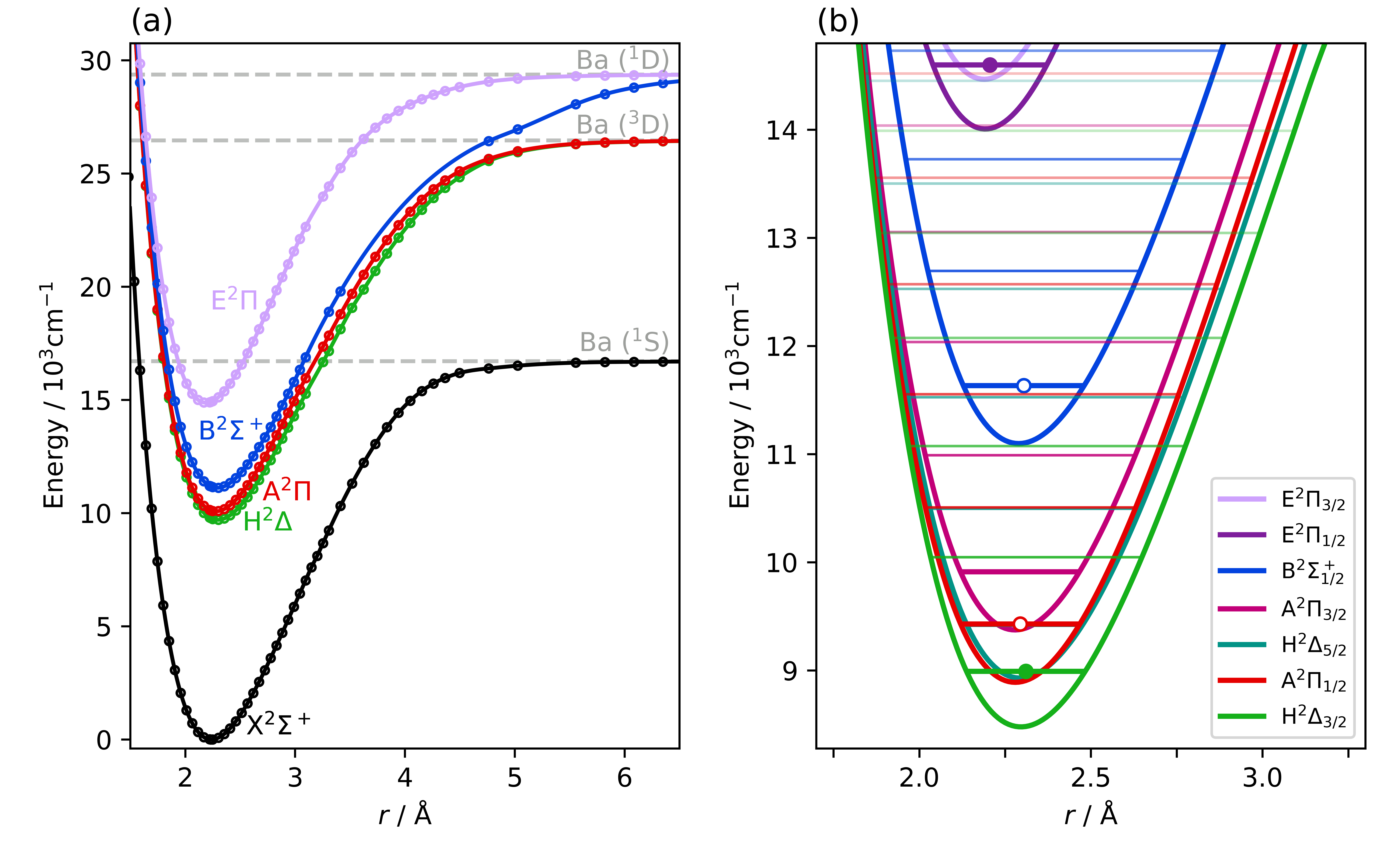}
\caption{(a) {\it Ab initio} potential energy curves of the BaH radical calculated by the MRCI method (with Davidson correction) using the ACV$Q$Z basis set on Ba.
The D$^2\Sigma^+$ potential, which dissociates into Ba(5$^{3}$D) atoms, has been removed from the figure. All hydrogen atoms correspond to the H(1\(^{2}\)S) state. The H$^2\Delta$ potential lies just below that for A$^2\Pi$ in the FC region.
(b) Details of the 5$d$ complex A$^2\Pi$, B$^2\Sigma^+$ and H$^2\Delta$ states including the spin-orbit splitting. The E$^2\Pi_{\Omega}$ ($\Omega$ = 1/2, 3/2) states are also plotted (all levels below 14800 cm$^{-1}$).
The lowest rotational energy for each vibronic level is plotted.
The vibrational spacings correspond to the {\it ab initio} values determined by \duo\ with the lowest vibrational level anchored in each state to experimental measurements (see text for more details).
The upper vibronic levels used in the competing cooling cycles are indicated by an unfilled roundel while the repumping transtion (see section \ref{Improv-cool}) is marked by the filled dots. The zero-energy in (b) is the energy of the lowest rovibrational level of the X$^2\Sigma^+$ state, around 580 cm$^{-1}$ (the zero-point energy) above the zero in (a). } \label{fig1}
\end{figure*}

The generated MRCI+Q (including the Davidson correction) potential energy curves for the electronic states with minima below the ground dissociation limit are shown in Fig.~\ref{fig1}(a).
The calculated equilibrium bond lengths $r_e$, electronic origins $T_e$ and well depths $D_e$ for all these states are documented in Table~\ref{tab:bah_ciq_eqval}.
The D$^2\Sigma^+$ state has been eliminated from the present analysis, even though it correlates to Ba atoms in the lowest excited state, because it lies significantly above the E$^2\Pi$ state in the Franck-Condon region. 

The equilibrium bond length in the resulting X$^{2}\Sigma^{+}$ state potential ($r_e$ = 2.23~\AA ) was a superior match to experiment~\cite{Ram2013} than  previous \cite{Allouche1992, Gao2014, Lane2015} theoretical studies save the CBS potential \cite{Moore2016} from Moore {\it et al.}
The biggest discrepancies are found for the states belonging to the 5$d$-complex, which is perhaps not surprising as the quadruple-zeta basis set here may be struggling to replicate fully the d-orbital on the Ba atom (the maximum angular momentum function present in the basis set is $l$ = 4), whereas the higher-lying E$^2\Pi$ state requires just a p-orbital and the present calculation is within 0.2~pm of the experimental \cite{Ram2013} value.
Furthermore, the calculated MRCI+Q vibrational spacings (Table~\ref{tab:bah_ciq_vibspace}) are in excellent agreement with experiment with all lying within 1\% of the measured values. The calculated ground state dissociation energy is 16708 cm$^{-1}$, in good agreement with the most recent {\it ab initio} results \cite{Moore2016}  \cite{Lesiuk2017}.

%% Vib only, spacings
\begin{table}
\center
\caption{Vibrational spacings from the initial \(J=0\) calculation performed using {\it ab initio} data and \duo\ \cite{Yurchenko2016} to compute the ro-vibronic levels. $\Delta$ is the difference between the calculated and observed energies.}
\label{tab:bah_ciq_vibspace}
\center
\footnotesize
\begin{tabular}{|lr|rrr|}
\noalign{\vskip 2mm}
\hline
state & \(v\)
	& \multicolumn{1}{c}{\(E_v\) / cm\(^{-1}\)}
	& \multicolumn{1}{c}{\(E_v-E_{v-1}\)}% / cm\(^{-1}\)}
	& \multicolumn{1}{c|}{\(\Delta\) / cm\(^{-1}\)}
	\rule{0pt}{2.5ex}\\[.8ex]\hline
\Xst
	& 0 &     0.00\(^\dagger\) && \rule{0pt}{2.5ex}\\ (\(a\))
	& 1 &  1130.48 & 1130.48 & -8.81 (-0.8\%)		\\
	& 2	&  2233.07 & 1102.58 & -7.73 (-0.7\%)	\\
	& 3 &  3207.83 & 1074.76 & -6.75 (-0.6\%)		\\[1ex]
\Hst
	& 0 &  9663.22 &&	\rule{0pt}{2ex}\\	(\(b\))
	& 1 & 10721.29 & 1058.07 &	\\
	& 2 & 11750.25 & 1028.96 &	\\[1ex]
\Ast
	& 0 & 10051.24 && \rule{0pt}{2ex}\\	(\(c\))
	& 1 & 11129.67 & 1078.42 & -2.96 (-0.3\%)	\\
	& 2 & 12177.86 & 1048.20 & -4.31 (-0.4\%)	\\[1ex]
\Bst
	& 0 & 11076.92 && \rule{0pt}{2ex}\\	(\(d\))
	& 1 & 12135.53 & 1058.61 & +0.57 (+0.1\%)	\\
	& 2 & 13165.98 & 1030.45 & +3.15 (+0.3\%)	\\
	& 3 & 14169.72 & 1003.74 & +7.04 (+0.7\%)	\\[1ex]
\Est
	& 0 & 14900.56 && \rule{0pt}{2ex}\\ (\(a\))
	& 1 & 16087.21 & 1186.66 & -3.92 (-0.3\%)	\\
	& 2 & 17241.16 & 1153.95 & -5.29 (-0.5\%)	\\[0.6ex]
\hline
\noalign{\vskip 2mm}
\multicolumn{5}{l}{\(\dagger\) ZPE = 575.35~\wn\ ($\Delta$ = -5.2~\wn, -0.9\%)\(^a\)}
\rule{0pt}{2ex}\\
\multicolumn{3}{l}{\(a\) vs. Ram and Bernath\cite{Ram2013}} &
	\multicolumn{2}{l}{\(c\) vs. Kopp \textit{et al.}\cite{Kopp1966-2}}	\\
\multicolumn{3}{l}{\(b\) vs. Bernard \textit{et al.}\cite{Bernard1989}} &
	\multicolumn{2}{l}{\(d\) vs. Appelblad \textit{et al.}\cite{Appelblad1985}}	\\
\end{tabular}
\end{table}

\subsection{Spin-orbit coupling}\label{CAtransition}
The program \texttt{DUO} was used \cite{Yurchenko2016} to solve the radial Schr{\"o}dinger equation for a system involving the sub-set of electronic states X\(^2\Sigma^+\), H\(^2\Delta\), A\(^2\Pi\), B\(^2\Sigma^+\) and E\(^2\Pi\).
For a \(^1\Sigma\) state this is given by

\begin{equation}
\frac{\hbar}{2\mu}\frac{d^2}{dr^2} \Psi_{vJ}(r) 
 + \left[V_{state} 
 + \frac{J(J+1)}{2\mu r^2}\right]\Psi_{vJ}(r)	
  =	E_{vJ}\Psi_{vJ}(r)
\end{equation}

However \duo\ is capable of treating the coupled state problem to account for spin-orbit and other interactions, allowing for a  more appropriate description of the ro-vibrational states in the cooling cycle.
When spin-orbit coupling is included the good electronic quantum number becomes $\Omega$.
While MOLPRO represents all calculations in the \(C_{2v}\) point group symmetry, \duo\ handles explicitly \(C_{\infty v}\) symmetry states and appropriate conversions \cite{Patrascu2014} are required to prepare MOLPRO output data for input into \duo: 

% Matrix Element Converters
\vspace{3cm}
{\flushleft
\begin{tabular}{lrl}
\multicolumn{3}{l}{\textbf{State Wavefunctions}}	\\[1ex]
	\(^2\Sigma^+\) states
		& {\small\(\ket{n\Sigma, 0, \pm\tfrac{1}{2}} = \)}
		& {\small\(\ket{n\sigma A_1, \pm\tfrac{1}{2}}\)}\\[.8ex]
	\(^2\Pi\) states
		& {\small\(\ket{n\Pi, \pm 1, \pm\tfrac{1}{2}} = \)}
		& {\small\(\mp \ket{n\pi B_1, \pm\tfrac{1}{2}}
		  - i\ket{n\pi B_2, \pm\tfrac{1}{2}}\)}\\[.8ex]
	\(^2\Delta\) states
		& {\small\(\ket{n\Delta, \pm 2, \pm\tfrac{1}{2}} = \)}
		& {\small\(\mp \ket{n\delta A_1, \pm\tfrac{1}{2}}
		  - i\ket{n\delta A_2, \pm\tfrac{1}{2}}\)}\\[1.5ex]
\noalign{\vskip 2mm}
\end{tabular}		  

\begin{tabular}{lrl}
\multicolumn{3}{l}{\textbf{Operators}}	\\[1ex]
	Dipole
		& \(\hat{\mu}_0 = \)
		& \(\hat{\mu}_z \)	\\[.8ex]
		& \(\hat{\mu}_\pm = \)
		& \(\mp \frac{1}{\sqrt{2}} \left( \hat{\mu}_x 
			\pm i\hat{\mu}_y \right) \)	\\[.9ex]
	Ladder
		& \(\hat{L}_\pm = \)
		& \(\hat{L}_x \pm i\hat{L}_y\)	\\[.9ex]
	Spin-Orbit
		& \(\hat{H}_{SO} = \)
		& \(\hat{H}_{SO,x} + \hat{H}_{SO,y} + \hat{H}_{SO,z}\)	\\[1.5ex]
\noalign{\vskip 2mm}
\end{tabular}		  

\begin{tabular}{lrl}
\multicolumn{3}{l}{\textbf{Energies}}	\\[1ex]
	\(^2\Sigma^+\) states
		& \(E_{n\Sigma^+} = \)
		& \(E_{n\sigma A_1}\)	\\[.8ex]
	\(^2\Pi\) states
		& \(E_{n\Pi} = \)
		& \(\tfrac{1}{2} \left(E_{n\pi B_1}
			+ E_{n\pi B_2}\right) \)	\\[.8ex]
	\(^2\Delta\) states
		& \(E_{n\Delta} = \)
		& \(\tfrac{1}{2} \left(E_{n\delta A_1}
			+ E_{n\delta A_2}\right) \)	\\[1.5ex]
\end{tabular}

\begin{tabular}{ll}
\multicolumn{2}{l}{\textbf{Spin-Orbit Matrix Elements}}	\\[1ex]
\({SO_{\Sigma-\Pi}}\)\\
\scriptsize{\(\mel{^2\Sigma^+,0,-\tfrac{1}{2}}{\hat{H}_{SO}}{^2\Pi,-1,+\tfrac{1}{2}} =\)}\\
\multicolumn{2}{l}{
	\quad \scriptsize{\(\frac{1}{\sqrt{2}}
	\left(\mel{\sigma A_1,-\tfrac{1}{2}}{\hat{H}_{SO,y}}{\pi B_1,+\tfrac{1}{2}}
	-i\mel{\sigma A_1,-\tfrac{1}{2}}{\hat{H}_{SO,x}}{\pi B_2,+\tfrac{1}{2}}\right)\)}}	\\[1.2ex]
\({SO_{\Pi-\Pi}}\)\\
\scriptsize{\(\mel{^2\Pi,+1,+\tfrac{1}{2}}{\hat{H}_{SO}}{^2\Pi,+1,+\tfrac{1}{2}} =\)}
&\scriptsize{\(i\mel{\pi B_1,+\tfrac{1}{2}}{\hat{H}_{SO,z}}{\pi B_2,+\tfrac{1}{2}}\)}	\\[1.2ex]
\({SO_{\Delta-\Pi}}\) \\
\scriptsize{\(\mel{^2\Delta,+2,-\tfrac{1}{2}}{\hat{H}_{SO}}{^2\Pi,+1,+\tfrac{1}{2}} =\)}\\
\multicolumn{2}{l}{\quad
	\scriptsize{\(\frac{1}{2}\left(
		\mel{\delta A_1,-\tfrac{1}{2}}{\hat{H}_{SO,y}}{\pi B_1,+\tfrac{1}{2}}
		+i\mel{\delta A_1,-\tfrac{1}{2}}{\hat{H}_{SO,x}}{\pi B_2,+\tfrac{1}{2}}\right.\)}}\\
\multicolumn{2}{l}{\quad\quad
	\scriptsize{\(\left.
		+i\mel{\pi B_1,-\tfrac{1}{2}}{\hat{H}_{SO,x}}{\delta A_2,+\tfrac{1}{2}}
		-\mel{\pi B_2,-\tfrac{1}{2}}{\hat{H}_{SO,y}}{\delta A_2,+\tfrac{1}{2}}
		\right)\)}}	\\[1.2ex]
\({SO_{\Delta-\Delta}}\)\\
\scriptsize{\(\mel{^2\Delta,+2,+\tfrac{1}{2}}{\hat{H}_{SO}}{^2\Delta,+2,+\tfrac{1}{2}} =\)}\\
& \scriptsize{\(i\mel{\delta A_1,+\tfrac{1}{2}}{\hat{H}_{SO,z}}{\delta A_2,+\tfrac{1}{2}}\)}
\\[.8ex]
\end{tabular}}

\vspace{3mm}
\noindent where $\ket{state, \Lambda, \Sigma}$ are the \(C_{\infty v}\) electronic wavefunctions \cite{Yurchenko2016} \cite{Patrascu2014} and $\sigma$, $\pi$ and $\delta$ are molecular orbitals. \duo\ requires both the spin-orbit coupling functions and the ladder coupling functions for the generation of energy levels with inclusion of spin orbit effects in the ro-vibrational eigenstates.
At points in the calculation the phase of a given wavefunction may invert, leading to a notable discontinuity in the function.
To ensure consistency the phase at the \(r_e\) is taken as the reference and any change in sign between two ab initio points is regarded as unphysical and reversed to maintain a smooth (slow) change in value. Additionally, where any state with \(\Lambda > 0\) is involved, the \(C_{\infty v}\) function will require contributions from multiple \(C_{2v}\) functions, which may in turn be sourced from  \(C_{2v}\) roots that change with internuclear separation.
For example, the \Hst\ and \Bst may readily switch in their \(nA_1\) representations.

There are currently three experimental values available for the BaH states of interest (Table~\ref{tab:bah_exp_data}) that are particularly sensitive to spin-orbit effects, namely the lifetime of the B$^2\Sigma^+_{1/2}$ state and the spin-orbit splitting in both the $^{2}\Pi$ states.
When the standard spin-orbit calculation routine in \molpro\ was adopted, the result for the lifetime was acceptable (though rather short) but the spin-orbit splittings were poor: for example, the splitting in the E$^{2}\Pi$ state was calculated as 296 cm$^{-1}$, just 64\% of its true value. The origin of the problem appeared to be the presence of an ECP to describe the barium atom. Unfortunately, if the alternative spin-orbit routine in MOLPRO (ECPLS) is used in its place the contributions from other atoms to the spin-orbit matrix elements are ignored. However, the molecule of interest is a diatomic and the lack of orbital angular momentum contributed from the hydrogen (\(^2\)S) should make this a relatively trivial concern. In addition, the lower lying electronic states of BaH \cite{Moore2016} have considerable ionic character, so effectively the hydrogen exists in the form of H$^{-}$ ($^{1}$S) and there is no spin-orbit contribution. 
 The calculated spectroscopic constants for all states with the spin-orbit corrections are presented in Table~\ref{tab:bah_SO_spectra}.
The calculated E$^{2}\Pi$ state splitting was now 434 cm$^{-1}$ which is just 30 cm$^{-1}$ smaller than the experimental \cite{Ram2013} value and the agreement in the H$^{2}\Delta$ state is excellent, justifying the approximation used.

%\begin{sidewaystable}
\begin{table}
\caption{\label{tab:bah_SO_spectra}Molecular constants (in cm$^{-1}$) for the
electronic states of BaH with minima below the first dissociation limit. $A_v$ is the (vibrational level dependent) spin-orbit constant.}
\footnotesize\setlength\tabcolsep{2pt}\center\begin{tabular}{lllllll}
\hline
$state$ & $v$ & $T_v$\textsuperscript{a} & $A_v$ & $B_v$ & $D_v\times10^4$ & Refs.
\rule{0pt}{2ex}\\\hline
\noalign{\vskip 1.1mm}
X$^2\Sigma^+$
  & 0 & 0.0 && {\color{blue}3.3274} & {\color{blue}1.116} & \textsuperscript{b}\\
  &   &     && 3.3495907(28) & 1.127057(64) & \textsuperscript{c}\\
  &   &     && 3.34986(5) & 1.1267(7) & \textsuperscript{e}\\
  & 1 & {\color{blue}1130.52} && {\color{blue}3.2633} & {\color{blue}1.111} & \textsuperscript{b}\\
  &   & 1139.289606(95) && 3.2838078(27) & 1.124169(66) & \textsuperscript{c}\\  
  &   & 1139.318(12)     && 3.2838(1)     & 1.116(3)     & \textsuperscript{e}\\  
  & 2 & {\color{blue}2233.15} && {\color{blue}3.1988} & {\color{blue}1.107} & \textsuperscript{b}\\
  &   & 2249.60618(14) && 3.2179014(30) & 1.120664(86) & \textsuperscript{c}\\
  &   & 2249.638(12)   && 3.2180(2)     & 1.116(4)     & \textsuperscript{e}\\  
  & 3 & {\color{blue}3307.97} && {\color{blue}3.1342} & {\color{blue}1.102} & \textsuperscript{b}\\
  &   & 3331.11924(19) && 3.1518703(34) & 1.117085(92) & \textsuperscript{c}\\
H$^2\Delta$
  & 0 & {\color{blue}9626.8} & {\color{blue}210.1} & {\color{blue}3.0935} & {\color{blue}0.973} &\textsuperscript{b}\\
  &   &  9207.491 & 217.298(86) & 3.11894(23) & 0.8947(67) & \textsuperscript{f, g}\\
  & 1 & {\color{blue}10685.4} & {\color{blue}212.0} & {\color{blue}3.0316} & {\color{blue}0.917} &\textsuperscript{b}\\
  &   & 10275.79 & 221.29(38) & 3.05687(10)  & 0.89       & \textsuperscript{f, g}\\
A$^2\Pi$
  & 0 & {\color{blue}10016.0} & {\color{blue}495.3} & {\color{blue}3.2432} & {\color{blue}1.151} &\textsuperscript{b}\\
  &   &  9669.623(23) & 482.51(2) & 3.2613(3) & 1.266(11) & \textsuperscript{e}\\
  & 1 & {\color{blue}11093.5} & {\color{blue}495.9} & {\color{blue}3.1690} & {\color{blue}1.195} &\textsuperscript{b}\\
  &   & 10751.008(50) & 485.34(5) & 3.1864(5) & 1.244(10) & \textsuperscript{e}\\
  & 2 & {\color{blue}12140.9} & {\color{blue}496.0} & {\color{blue}3.0953} & {\color{blue}1.223} &\textsuperscript{b}\\  
  &   & 11803.513(51) & 491.57(5) & 3.1059(4) & 1.143(11) & \textsuperscript{e}\\
B$^2\Sigma^+$
  & 0 & {\color{blue}11167.1} && {\color{blue}3.2307} & {\color{blue}1.126} &\textsuperscript{b}\\
  &   & 11633.1755(15) && 3.233484(8) & 1.15702(11) & \textsuperscript{d}\\
  & 1 & {\color{blue}12227.6} && {\color{blue}3.1633} & {\color{blue}1.168} &\textsuperscript{b}\\
  &   & 12691.2104(19) && 3.162726(13) & 1.15406(22) & \textsuperscript{d}\\
  & 2 & {\color{blue}13259.7} && {\color{blue}3.0986} & {\color{blue}1.330} &\textsuperscript{b}\\
  &   & 13718.5118(23) && 3.091898(34) & 1.15217(85) & \textsuperscript{d}\\
  & 3 & {\color{blue}14264.1} && {\color{blue}3.0205} & {\color{blue}1.078} &\textsuperscript{b}\\
  &   & 14715.2147(34) && 3.021146(57) & 1.1582(17)  & \textsuperscript{d}\\
E$^2\Pi$
  & 0 & {\color{blue}14911.5} & {\color{blue}429.4} & {\color{blue}3.4873} & {\color{blue}0.980} &\textsuperscript{b}\\
  &   & 14856.63369(38) & 461.85585(79) & 3.4868233(40) & 1.167801(85) &\textsuperscript{c}\\
  &   & 14859.889(6)    & 462.3046(80)  & 3.48510(12)   & 1.1599(40)   &\textsuperscript{f}\\
  & 1 & {\color{blue}16098.2} & {\color{blue}436.0} & {\color{blue}3.4176} & {\color{blue}1.040} &\textsuperscript{b}\\
  &   & 16047.20902(65) & 469.9424(13) & 3.414457(14) & 1.19689(98) &\textsuperscript{c}\\
\hline
\noalign{\vskip 1.2mm}
\multicolumn{4}{l}{\textsuperscript{a} Related to the $X^2\Sigma^+, v=0$ level.}
&	\multicolumn{3}{l}{\textsuperscript{e} Kopp et al.\cite{Kopp1966-2}.}\\
\multicolumn{4}{l}{\textsuperscript{b} This work \emph{ab initio} results.}
&	\multicolumn{3}{l}{\textsuperscript{f} Fabre et al.\cite{Fabre1987}.}\\
\multicolumn{4}{l}{\textsuperscript{c} Ram and Bernath \cite{Ram2013}.}
&	\multicolumn{3}{l}{\textsuperscript{g} Bernard et al. used for \(A_v\)\cite{Bernard1989}.}\\
\multicolumn{4}{l}{\textsuperscript{d} B-X experimental results \cite{Appelblad1985}}\\
%\multicolumn{7}{l}{\textsuperscript{e} Kopp et al.\cite{Kopp1966-2}.}\\
%\multicolumn{7}{l}{\textsuperscript{f} Fabre et al.\cite{Fabre1987}.}\\
%\multicolumn{7}{l}{\textsuperscript{g} Bernard et al. used for \(A_v\)\cite{Bernard1989}.}\\
\end{tabular}
\end{table}
%\end{sidewaystable}

\begin{figure}
\center
\includegraphics[width=90mm]{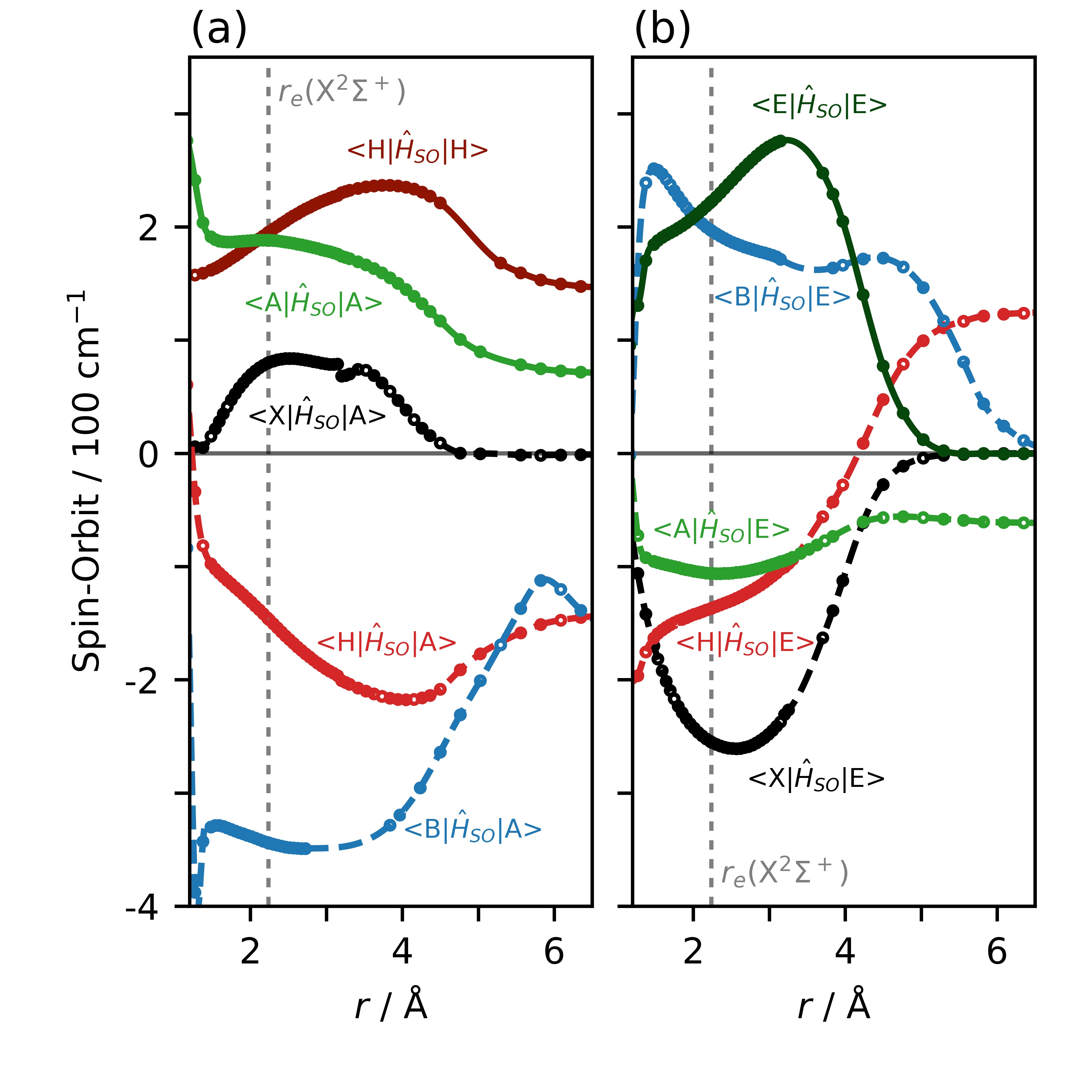}
\caption{Generated spin-orbit coupling functions as a function of interatomic separation for (a) the 5$d$-complex states and (b) the E$^2\Pi$ state. { \(\bra{H}\)}$H_{SO}${ \(\ket{A}\)} $\equiv$
{\scriptsize \(\bra{\Hst, +2, -\tfrac{1}{2}}\)}$H_{SO}${\scriptsize \(\ket{\Ast, +1, +\tfrac{1}{2}}\)} etc (see text). The minimum of the X state potential is marked with the dashed grey lines.
Spin-orbit coupling matrix elements for BaH computed (dots) with MRCI wavefunctions using the  ACV$Q$Z basis set. The interpolated region on matrix elements involving the B$^2\Sigma^+$ state (dashed lines) is due to the switch of electron orbital occupancies to that of the D$^2\Sigma^+$ state for the lowest excited $^2\Sigma^+$ potential \cite{Moore2016}. Note this is far outside the FC region of concern here, as are the minor discontinuities in the curves.}
\label{fig:SOfuns}
\end{figure}

\begin{figure}
\center
\includegraphics[width=90mm]{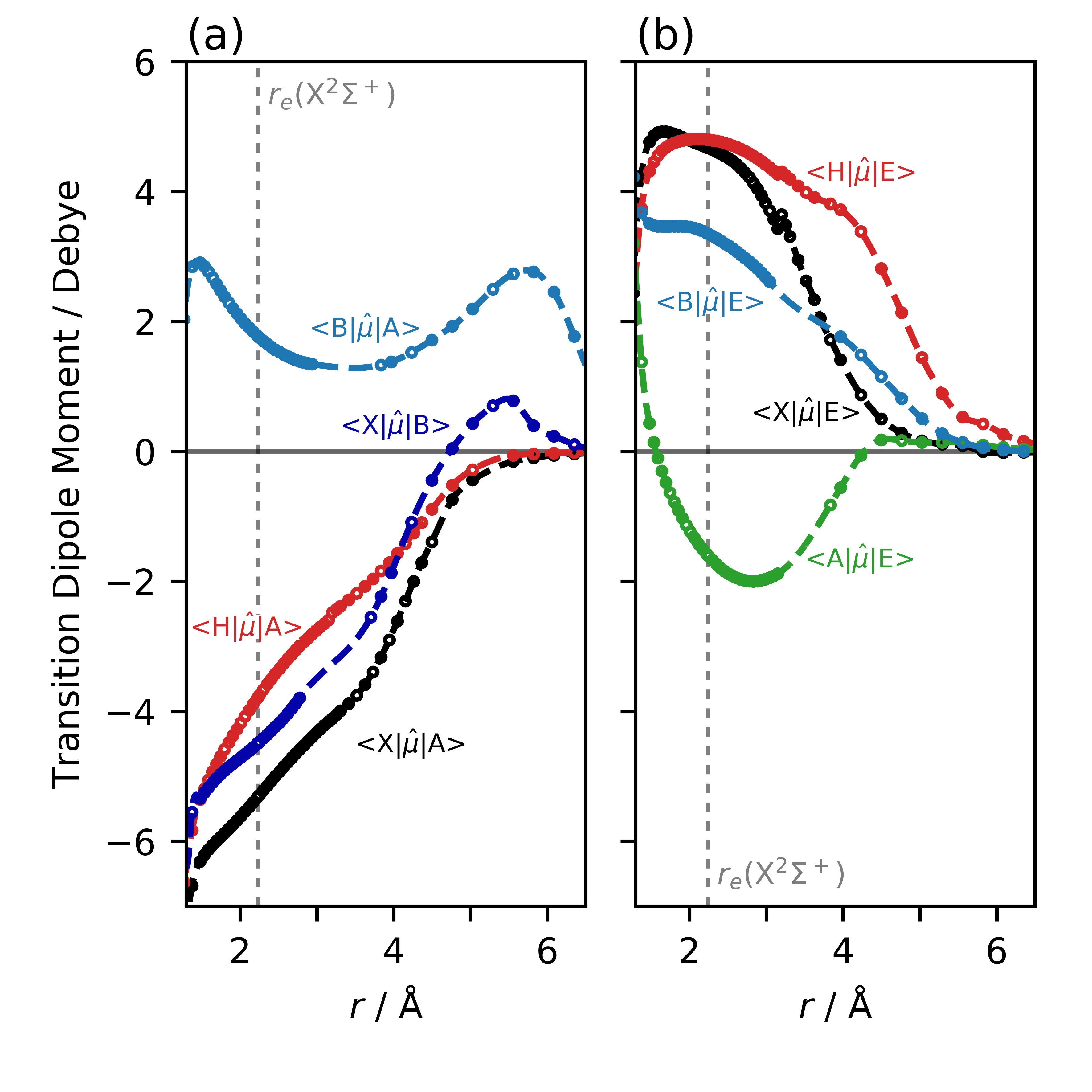}
\caption{{\it Ab initio} electric dipole (E1) transition dipole moments (TDMs) as a function of interatomic separation of (a) the bound-bound electronic transitions in the BaH radical involving the 5$d$-complex states as the upper electronic level.
This not only covers all the practical laser cooling transitions but also all the radiative decay pathways.
Moments determined by the MRCI method using the ACV$Q$Z basis set (dots). Note { \(\bra{X}\)}$\mu${\(\ket{H}\)} = 
{ \(\bra{H}\)}$\mu${\(\ket{B}\)} = 0 because of the selection rules for E1 transitions in the absence of any spin-orbit mixing.
(b) {\it Ab initio} E1 TDMs where E$^2\Pi$ is the upper electronic state. All transitions are forbidden at the atomic asymptotes because of the Laporte \cite{Laporte1925} selection rule.}
\label{fig:cinf_TDM_all}
\end{figure}

\begin{figure}
\center
\includegraphics[width=90mm]{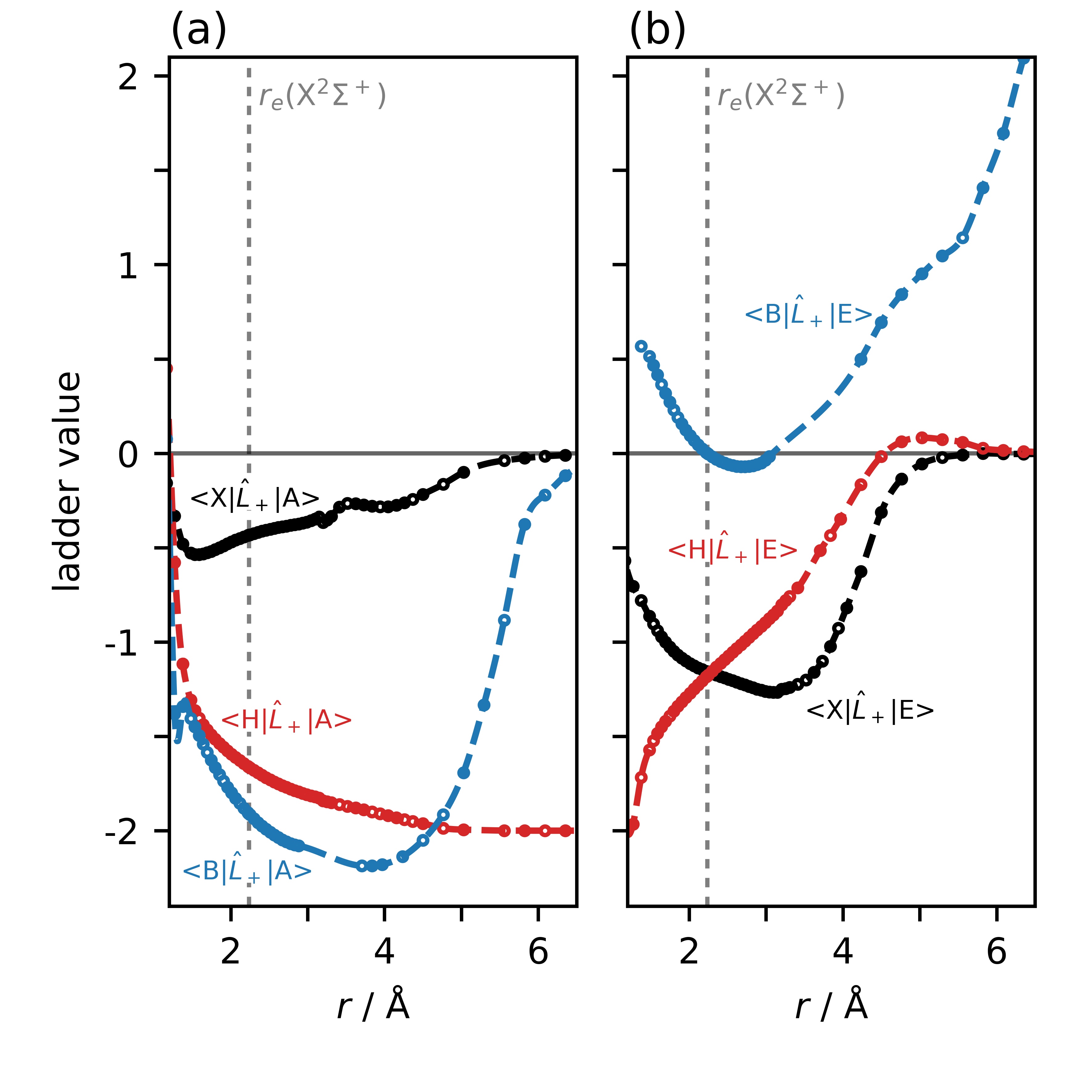}
\caption{Calculated ladder matrix element functions (dots) for all the relevant states as a function of interatomic separation for (a) the A$^2\Pi$ state and (b) the E$^2\Pi$ state. For non-zero matrix elements, $\Delta \Lambda$ = $\pm$1.}
\label{fig:cinf_ladder_all}
\end{figure}

Fig.~\ref{fig:SOfuns} presents the spin-orbit functions determined in this five state calculation.
It can be readily seen that some discontinuities still persist in the functions, which often correspond to crossings or avoided-crossings in the parents potentials.
Ideally such discontinuities might be smoothed through use of a polynomial-type fit.
Similar discontinuities can also be seen in the dipole functions for the five state calculation (Fig.~\ref{fig:cinf_TDM_all}).
However, they lie well outside the Franck-Condon region which is the focus of the transitions considered in this particular study.
The calculated angular momentum couplings $L_x$ and $L_y$ matrix elements are shown in Fig.~\ref{fig:cinf_ladder_all}.

% Spin-orbit Correction
%\begin{figure}
%\center
%\includegraphics[width=0.9\linewidth]{bah_qz_8331_ecpls_XsoA_process}
%\caption{Conversion of \(C_{2v}\) spin-orbit functions to \(C_{\infty v}\).
%(\textbf{top}) \(C_{2v}\) display unphysical ``sign-flipping'' at certain points in the function.
%(\textbf{middle}) Correction of this behaviour yields two analogous components of the \(C_{\infty v}\) function which (\textbf{bottom}) are readily combined.}
%\label{fig:SOconv}
%\end{figure}
%

% Spin-orbit Functions
%

% Transition Dipole functions
%

% Ladder functions
%

% re matrix elements
\begin{table}
\center
\footnotesize
\caption{Values for the evaluated matrix elements at the equilibrium distance of the \Xst\ state (2.232~\AA~\cite{Ram2013}).
Spin-orbit coupling and ladder matrix elements in cm$^{-1}$.}
\label{tab:matelem_values}
\begin{tabular}{|llrrr|}
\noalign{\vskip 5mm}
\hline
\hspace{2.5mm}State 1 & \hspace{2.5mm}State 2 \rule{0pt}{2.5ex} &
	Dipole/D & Spin-Orbit  &  Ladder	\\[.8ex]\hline
%
% XX
{\scriptsize \(\bra{\Xst, 0+\tfrac{1}{2}}\)} 
& {\scriptsize \(\ket{\Xst, 0+\tfrac{1}{2}}\)} \rule{0pt}{2.5ex}
	&  -3.3211  &  --  &  --  \\[.8ex] \hline
%%
% HH
{\scriptsize \(\bra{\Hst, +2, +\tfrac{1}{2}}\)}
&{\scriptsize \(\ket{\Hst, +2, +\tfrac{1}{2}}\)} \rule{0pt}{2.5ex}
	&  -6.3648  &   194.98  &  --  \\[.8ex] \hline
%%
% XA
{\scriptsize \(\bra{\Xst, 0, -\tfrac{1}{2}}\)}
&{\scriptsize \(\ket{\Ast, -1, +\tfrac{1}{2}}\)} \rule{0pt}{2.5ex}
	&  -5.3128  &    79.93  & -0.434  \\[.5ex]
% HA
{\scriptsize \(\bra{\Hst, +2, -\tfrac{1}{2}}\)}
&{\scriptsize \(\ket{\Ast, +1, +\tfrac{1}{2}}\)} \rule{0pt}{2.5ex}
	&  -3.7807  &  -145.91  &  -1.661  \\[.5ex]
% AA
{\scriptsize \(\bra{\Ast, +1, +\tfrac{1}{2}}\)}
&{\scriptsize \(\ket{\Ast, +1, +\tfrac{1}{2}}\)} \rule{0pt}{2.5ex}
	&  -2.3139  &   188.23  &  --  \\[.8ex] \hline
%%
% XB
{\scriptsize \(\bra{\Xst, 0, +\tfrac{1}{2}}\)}
&{\scriptsize \(\ket{\Bst, 0, +\tfrac{1}{2}}\)} \rule{0pt}{2.5ex}
	&  -4.4898  &  --  &  --  \\[.5ex]
% BA
{\scriptsize \(\bra{\Bst, 0, -\tfrac{1}{2}}\)}
&{\scriptsize \(\ket{\Ast, -1, +\tfrac{1}{2}}\)} \rule{0pt}{2.5ex}
	&   1.7716  &  -343.86  &  -1.908  \\[.5ex]
% BB
{\scriptsize \(\bra{\Bst, 0, +\tfrac{1}{2}}\)}
&{\scriptsize \(\ket{\Bst, 0, +\tfrac{1}{2}}\)} \rule{0pt}{2.5ex}
	&  -1.7097  &  --  &  --  \\[.8ex] \hline
%%
% XE
{\scriptsize \(\bra{\Xst, 0, -\tfrac{1}{2}}\)}
&{\scriptsize \(\ket{\Est, -1, +\tfrac{1}{2}}\)} \rule{0pt}{2.5ex}
	&   4.6777  &  -254.75  &  -1.157  \\[.5ex]
% HE
{\scriptsize \(\bra{\Hst, +2, -\tfrac{1}{2}}\)}
&{\scriptsize \(\ket{\Est, +1, +\tfrac{1}{2}}\)} \rule{0pt}{2.5ex}
	&   4.8019  &  -136.92  &  -1.181  \\[.5ex]
% AE
{\scriptsize \(\bra{\Ast, +1, +\tfrac{1}{2}}\)}
&{\scriptsize \(\ket{\Est, +1, +\tfrac{1}{2}}\)} \rule{0pt}{2.5ex}
	&  -1.5892  &  -105.91  &  --  \\[.5ex]
% BE
{\scriptsize \(\bra{\Bst, 0, -\tfrac{1}{2}}\)}
&{\scriptsize \(\ket{\Est, -1, +\tfrac{1}{2}}\)} \rule{0pt}{2.5ex}
	&   3.3588  &   197.01  &   0.002  \\[.5ex]
% EE
{\scriptsize \(\bra{\Est, +1, +\tfrac{1}{2}}\)}
&{\scriptsize \(\ket{\Est, +1, +\tfrac{1}{2}}\)} \rule{0pt}{2.5ex}
	&  -9.0361  &   222.11  &  --  \\[.8ex] \hline
\end{tabular}
\end{table}

The calculated spin-orbit coupling for the 5$d$-complex states H$^2\Delta$ and A$^2\Pi$ are both within 4\% of spectroscopic measurements.
In the case of A$^2\Pi$ this is the experimental result from Kopp {\it et al.} \cite{Kopp1966-2} rather than the most recent \cite{Bernard1989} value.
Table~\ref{tab:matelem_values} presents the coupling matrix elements calculated at $r_e$.
The values of 
{\scriptsize \(\bra{\Hst, +2, -\tfrac{1}{2}}\)}$H_{SO}${\scriptsize \(\ket{\Ast, +1, +\tfrac{1}{2}}\)} and 
 {\scriptsize \(\bra{\Ast, -1, +\tfrac{1}{2}}\)}$H_{SO}${\scriptsize \(\ket{\Bst, 0, -\tfrac{1}{2}}\)} (\mbox{-145.91}~cm$^{-1}$ and \mbox{-343.86}~cm$^{-1}$) are consistent with the experimental values \cite{Bernard1987, Bernard1989} derived from Bernard {\it et al.}, though the signs are reversed.
This change of sign has no effect on the computed energy values. The final SO potential curves are shown in Fig.~\ref{fig1}(b) along with the corresponding $N$ = 0 rovibronic energy levels as calculated using \texttt{DUO}.
 
The calculated minima (Table~\ref{tab:bah_ciq_eqval}) are within 1 pm of the experimental values except for the A$^2\Pi$ and B$^2\Sigma^+$ states that belong to the 5$d$ complex.
Even the $T_e$ values computed for the 5$d$ complex (B$^2\Sigma^+$, A$^2\Pi$ and H$^2\Delta$) states are notably less accurate than the other electronic states studied for the reasons discussed earlier in Section~\ref{PECurves} and including the SO coupling does not dramatically improve that agreement (Table~\ref{tab:bah_SO_spectra}).
However, the computed SO splittings and ro-vibrational spacings are excellent but clearly properties that are particularly sensitive to either $r_e$ or $T_e$, such as the Einstein A-coefficients, need to be corrected.

%----------------------------------------------------------------
\section{
\label{sec:BaHCool}
Laser cooling BaH}
%----------------------------------------------------------------

A successful laser cooling strategy for a molecule requires a strongly diagonal electronic transition that does not suffer significant parasitic losses, such as predissociation or decay to a dark state.
All the states discussed in this paper lie below the lowest dissociation limit so only radiative decay outside the cooling cycle is of concern.

\begin{figure}
\centering
\includegraphics[width=90mm]{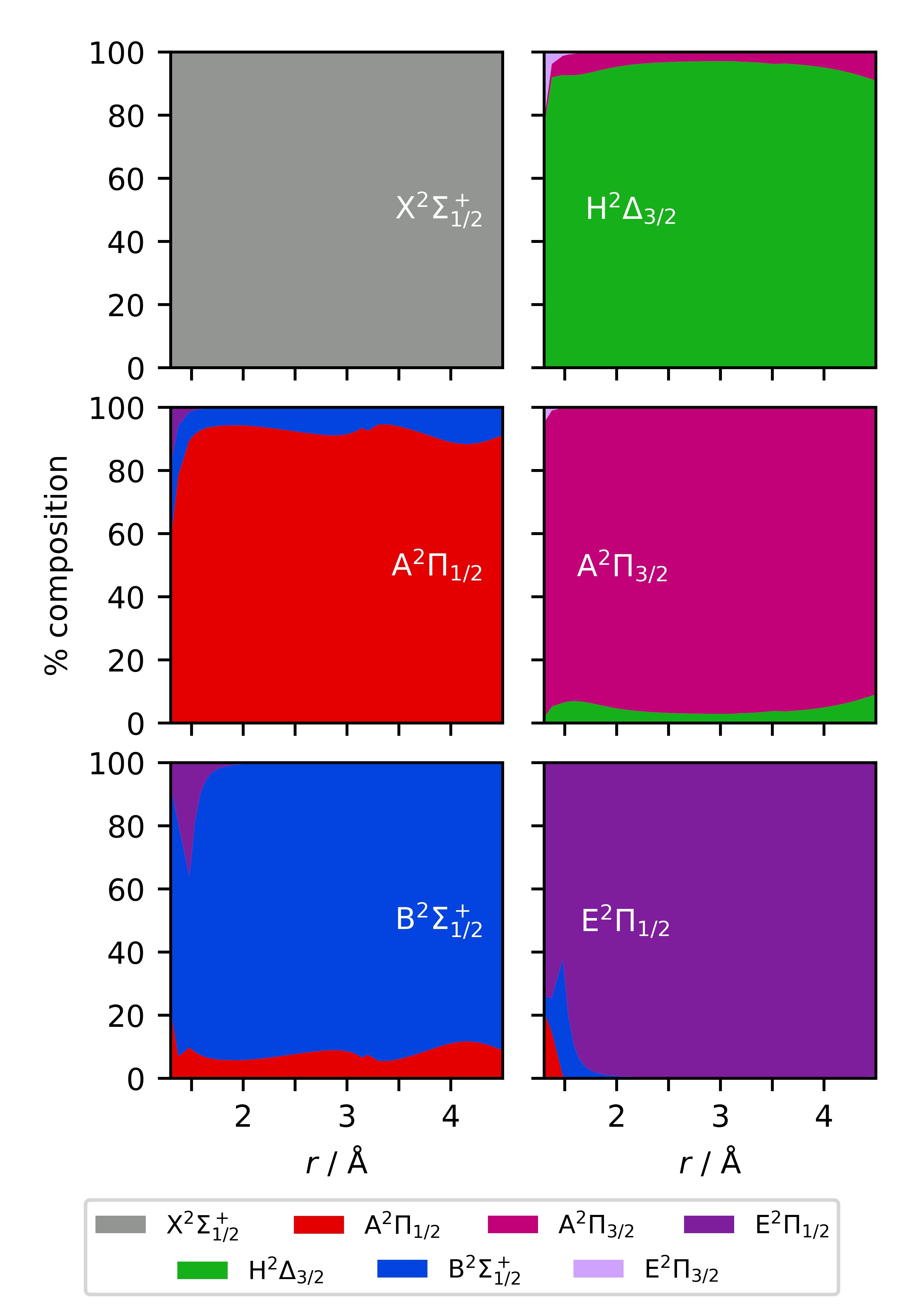}
\caption{Mixing of Hund's a states by the spin-orbit coupling for the $\Omega$ states of the 5$d$ complex (except H$^2\Delta_{5/2}$ which does not mix), E$^2\Pi$ and the ground X$^2\Sigma^+$.   } \label{fig:mixing}
\end{figure}

\subsection{Transition dipole moments}
\begin{figure}
\center
\includegraphics[width=90mm]{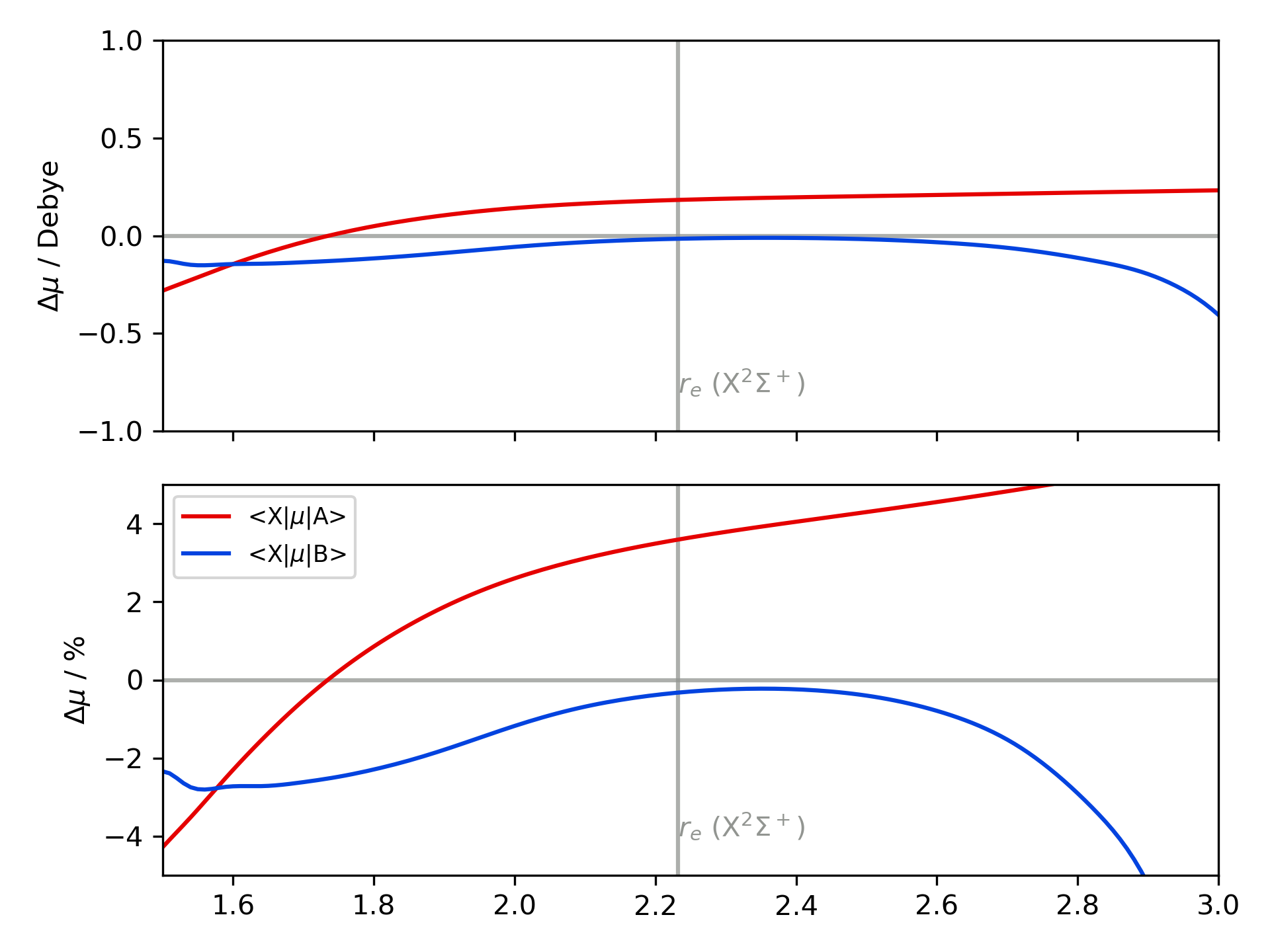}
\caption{The effect of SA-CASSCF wavefunctions on the final MRCI TDMs.
The curves present the differences in the {\it ab initio} TDMs when a CAS-5222/MRCI-5222 or a CAS-7442/MRCI-3221 calculation is performed.}
\label{fig:MRCI_TDMs}
\end{figure}

The transition dipole moments (TDMs) that are computed at the MRCI level are shown in Fig.~\ref{fig:cinf_TDM_all}.
This step is done within the spin-orbit code of MOLPRO and was performed to ensure a much higher accuracy than possible with CASSCF wavefunctions.
However, it was observed that the nature of the underlying CASSCF/MRCI wavefunctions strongly affected the final TDM values. In Fig.~\ref{fig:MRCI_TDMs}, the same basis set and active space (8$a_1$3$b_1$3$b_2$1$a_2$) is used in two different TDM calculations. The first uses a CAS-6222 calculation followed by an MRCI calculation MRCI-5222, while in the second case a CAS-7442 calculation was followed by MRCI-3221.
In both cases two \(^2B_1\) states featured at the MRCI level yet the {\it ab initio} A--X and \mbox{B--X} TDMs showed discrepancies between the calculated values of typically between 2- 5\%.
The result was a difference of as much as 10\% in the calculated lifetime of the A$^2\Pi$ state, while the change in B$^2\Sigma^+$ was less than 5\% for $v$ = 0 and 1. Clearly, the difficulty in determining an accurate Dipole Moment Function (DMF) is a major obstacle to quantitative calculations of the cooling dynamics. The CAS-7442/MRCI-3221 calculation was ultimately adopted because it correctly identifies 2\(^2A_1\) = 1\(^2A_2\) (the H$^2\Delta$ state) at $r_e$ and has a lower energy minimum for  X$^2\Sigma^+$.

These {\it ab initio} results show that the B$^2\Sigma^+$ -- X$^2\Sigma^+$ and A$^2\Pi$ -- X$^2\Sigma^+$ TDMs are large and almost identical across the Franck-Condon region associated with the ground vibrational wavefunction of the X$^2\Sigma^+_{1/2}$ state.
Also strong is the TDM connecting the A$^2\Pi$ and H$^2\Delta$ states but somewhat weaker than all these is the B$^2\Sigma^+$ -- A$^2\Pi$ moment. These latter TDMs are significant because they can disrupt the A$^2\Pi$ -- X$^2\Sigma^+$ and B$^2\Sigma^+$ -- X$^2\Sigma^+$ cooling cycles respectively.

The spin-orbit coupling modifies the strength of the radiative transitions between states by mixing the bare Hund's case (a) wavefunctions.
The relative case (a) components of the final $\Omega$ states from X$^2\Sigma^+_{1/2}$ to E$^2\Pi_{1/2}$  are shown in Fig.~\ref{fig:mixing}.
Significant mixing takes place amongst the $\Omega$-components of the 5$d$-complex and this will lead to intensity borrowing.
By contrast, the X$^2\Sigma^+_{1/2}$ state is barely changed by SO-coupling while the E$^2\Pi_{1/2}$ state has limited contributions from the lower states only at short range. 

% TDMs

The final rovibronic energy levels are then computed and the Einstein A-coefficients determined for all allowed transitions.
These latter values are then adjusted by replacing the {\it ab initio} transition frequencies  with the experimentally determined values.
For the B$^2\Sigma^+_{1/2}$-state the experimental data \cite{Appelblad1985} was taken from Appelblad {\it et al.} (the effect of this can be seen in Fig.~\ref{fig:Shift_effects}(b)) while those for the A$^2\Pi_{1/2, 3/2}$-states were published \cite{Kopp1966-2} by Kopp, Kronekvist and Guntsch.
The spectroscopic study by Ram and Bernath \cite{Ram2013} on the E$^2\Pi_{1/2}$ -- X$^2\Sigma^+_{1/2}$ transition provides the constants used in this work for both these states.
Finally, the spin-orbit coupling data \cite{Bernard1989} from Bernard {\it et al.} was used for H$^2\Delta$ in combination with the experimental H$^2\Delta_{5/2}$ $\nu^{\prime}$~=~0, $J$ = 5/2 value \cite{Fabre1987} from Verges and co-workers.

\subsection{The B$^2\Sigma^+_{1/2} -$ X$^2\Sigma^+_{1/2}$ and $A^2\Pi_{1/2} - X^2\Sigma^+_{1/2}$ transitions}

\begin{figure}
\center
\includegraphics[width=90mm]{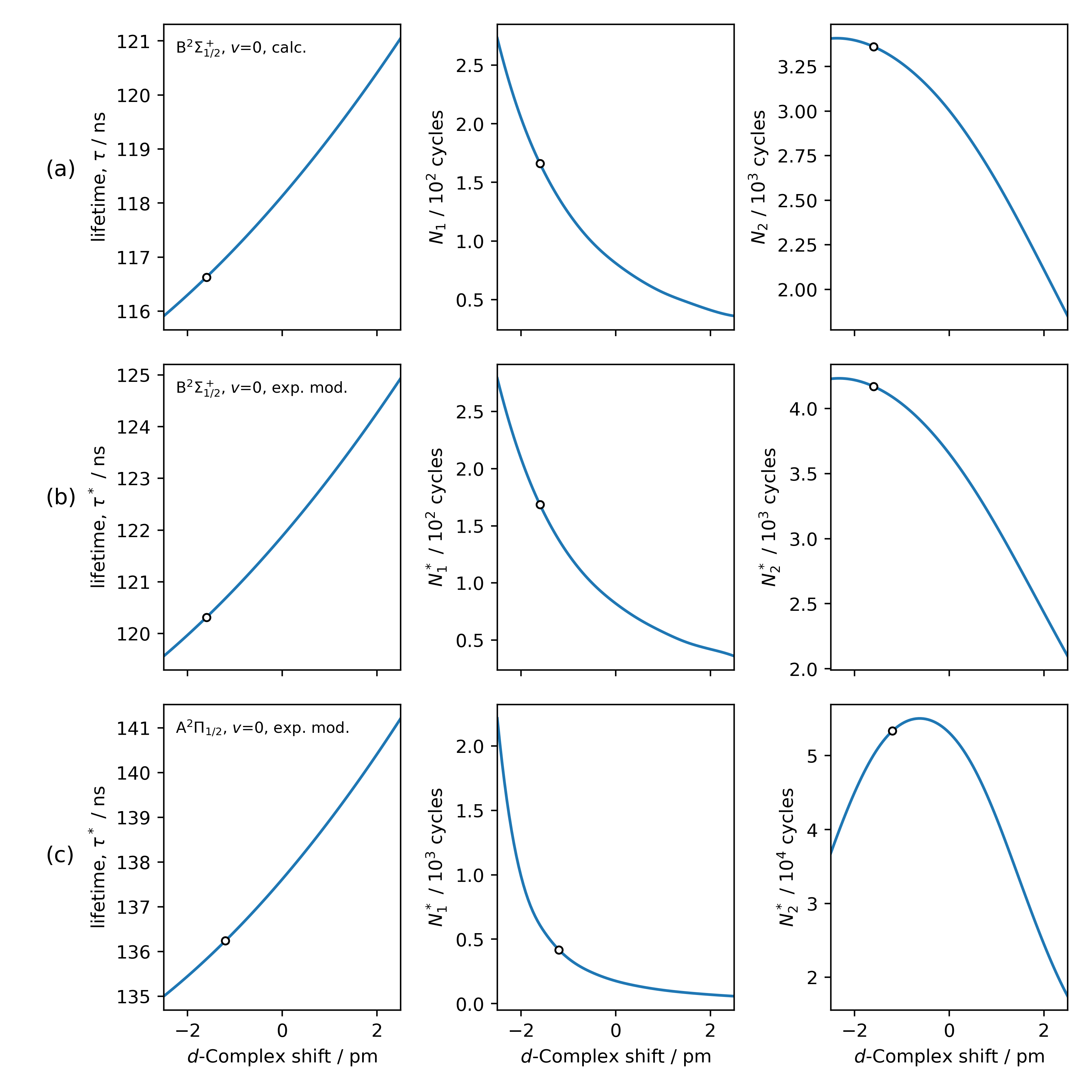}
\caption{Fundamental laser cooling parameters for BaH computed with MRCI+Q wavefunctions using the  ACV$Q$Z basis set.
The top row (a) presents the effect of small shifts in the $r_e$ value for the B$^2\Sigma^+_{1/2}$ state on the lifetime (left panel), the number of one colour cooling cycles $N_1$ (middle panel) and the two-colour cooling cycles $N_2$ (right panel).
In (b) experimental data is used to correct the energies of the ro-vibronic levels prior to calculation of the Einstein A-coefficients.
In (c) the experimentally corrected parameters for the A$^2\Pi_{1/2}$ state. Also marked (hollow dots) are the shifts in the {\it ab initio} bond length required to match the experimental difference $\Delta r_e$ between the upper state and X$^2\Sigma^+_{1/2}$.}
\label{fig:Shift_effects}
\end{figure}

\begin{figure*}
\centering
\includegraphics[width=140mm]{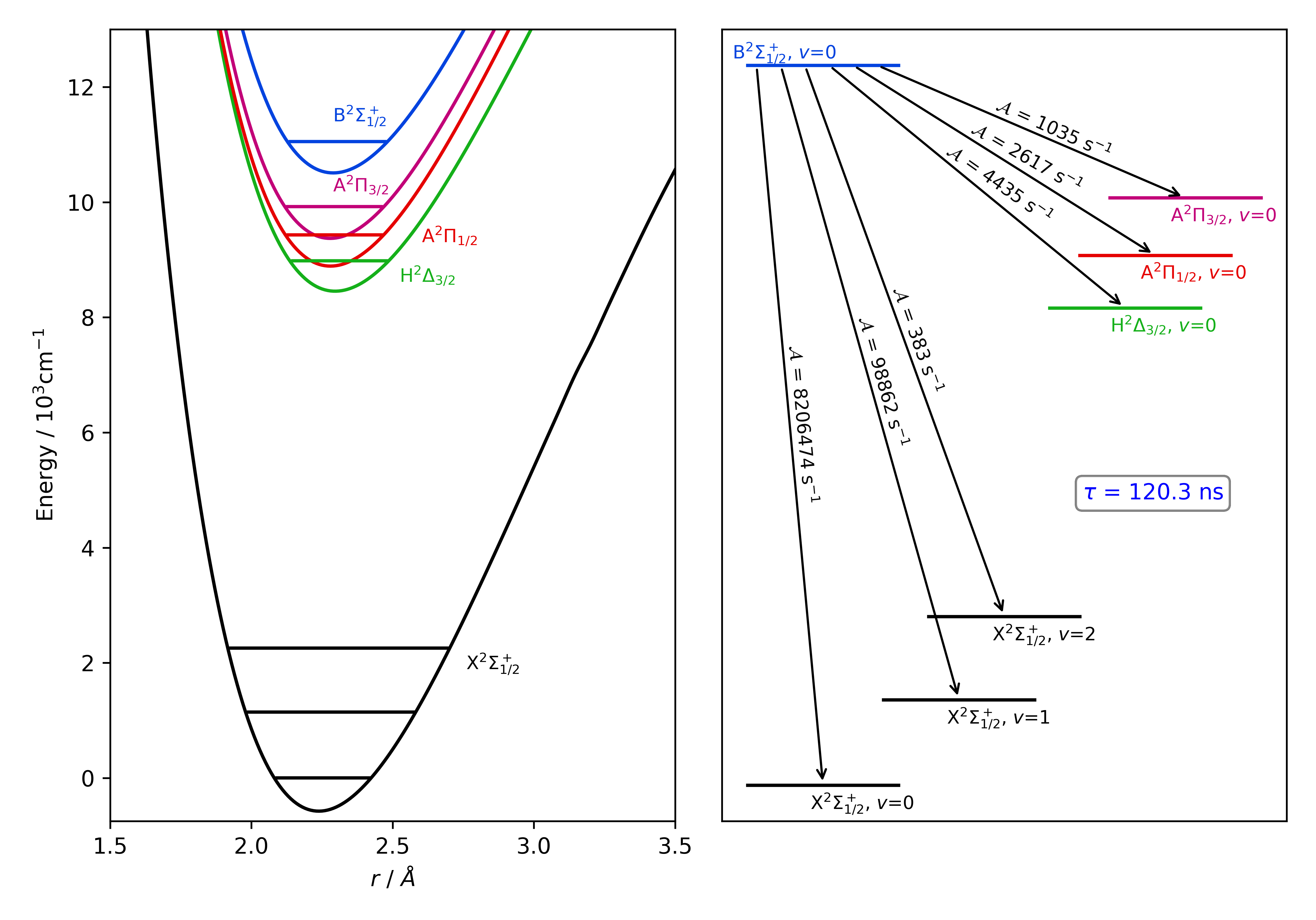}
\caption{The decay pathways from pumping the lowest vibrational level of the B$^2\Sigma^+$ state on the B$^2\Sigma^+ -$ X$^2\Sigma^+$ transition.
Decay levels within 1000 cm$^{-1}$ of the upper B$^2\Sigma^+$ v = 0 level are omitted from the figure on the right because the resulting A-coefficients ($\mathcal{A}$) are below 100 s$^{-1}$. }\label{fig6}
\end{figure*}

Perhaps the most significant effect of the mixing is a large A$^2\Pi$ contribution to the 
B\(^2\Sigma^+\) wavefunction.
Nominally, there can be no transition between the B$^2\Sigma^+$ state and the lower H$^2\Delta$ but by
borrowing intensity from the strong \(^2\Pi \rightarrow \) $^2\Delta$ transition, a significant decay pathway B\(^2\Sigma^+_{1/2} \rightarrow  \) H$^2\Delta_{3/2}$ opens up (Table~\ref{table6}).
However, discrepancies between the calculated vales of $T_e$ and $r_e$ can lead to errors in the decay rates and ultimately the excited state lifetimes.
The effect of shifting the value of $r_e$ for the calculated  B$^2\Sigma^+$ state on its lifetime is presented in Fig.~\ref{fig:Shift_effects}(a).
There is a significant but smooth dependence (left hand panel) on $r_e$ even over a range of just $\pm$ 2 pm. When considering FC-Factors it is the difference in $r_e$ between the lower and upper states that is especially important in determining their final magnitude.  
The observed experimental $\Delta r_e$ between the X$^2\Sigma^+$ and B$^2\Sigma^+$ states corresponds to a shift in the B$^2\Sigma^+$ potential minimum marked by the hollow dot in Fig.~\ref{fig:Shift_effects}(a).  To minimise the effect of errors in $r_e$, the calculated B$^2\Sigma^+$ state is shifted by this value prior to determination of the decay channels.
All the tabulated theoretical values are performed following this transformation and the equivalent shift for the A$^2\Pi_{1/2}$ state. 
The available decay paths and their relative strengths are shown in Fig.~\ref{fig6}.
The calculated lifetime of the B$^2\Sigma^+_{1/2}$ state of 120.3 ns is in good agreement with the lifetime of the $J$~=~11/2 level measured~\cite{Berg1997} by Kelly and co-workers. 

% TDMs

Unlike B$^2\Sigma^+_{1/2}$, there is no lifetime measurement currently available for A$^2\Pi_{1/2}$ state.
The calculated $v^\prime$ = 0 lifetime (Fig.~\ref{fig:Shift_effects}(c)) is 136.2 ns, slightly longer than the corresponding vibrational level in   
B$^2\Sigma^+$.
The difference is consistent both with the change in the $\omega^3$ term in the lifetime formula \cite{Hansson2005} on moving from the B$^2\Sigma^+$ to the lower lying A$^2\Pi$ state and that the A - X transition dipole moments is slightly larger across the ground state $v^{\prime\prime}$ = 0 wavefunction. In addition, comparing a  \texttt{DUO} calculation that includes the E$^2\Pi$ state with another that does not reveals that  mixing with the E$^2\Pi_{1/2}$ state at shorter range lowered the lifetime of the A$^2\Pi_{1/2}$ state by around 5 ns. The apparently much shorter lifetime predicted \cite{Gao2014} by Gao {\it et al.} can also be explained as simply the result of unresolved F$_1$ and F$_2$ components for each rotational level in the $^2\Pi$ state, effectively leading to a double counting of ro-vibronic levels (a consequence of not including the electron spin because of the restriction \cite{LeRoy2017} to singlet states in the \texttt{LEVEL} code). 

\subsection{Comparison of laser cooling strategies in BaH}\label{Compare-cool}

The leading candidates for cooling transitions involve the lowest rotational levels in the A\(^2\Pi\) and B\(^2\Sigma^+\) states optically driven from the ground X\(^2\Sigma^+\) state.
They have almost identical TDMs and by using two lasers, to repopulate them from the lowest pair of vibrational levels in the X\(^2\Sigma^+\) state, both transitions have very similar efficiencies.
Crucially, however, the A -- X cooling has a single loss channel involving another electronic state (A -- H) while the  B -- X transition has three (stronger) decay routes as shown in Table~\ref{table6}. 

\begin{table}
\centering
\caption{Decay pathways in the lowest vibronic levels of the A$^2\Pi_{1/2}$, B$^2\Sigma^+_{1/2}$ and H$^2\Delta_{3/2}$ states of the BaH molecule. $\mathcal{A}$ is the Einstein A-coefficient for each transition, $\mathcal{R}$atio is the value of $\mathcal{R}_{ij}$ in percentage terms and \(v''\) is the final vibrational level.}
                                           
\label{table6}
{\footnotesize
\begin{tabular}{|l|l|llrr|}
\noalign{\vskip 5mm}
\hline
Excited State	&	\(\tau\) / ns		\rule{0pt}{2.5ex}	& \multicolumn{4}{|c|}{Decay pathways}		\\
(lower level)	&&	Final State	& \(v''\)	&	\(\mathcal{A}\) / s\(^{-1}\)	& $\mathcal{R}$atio\hspace{2mm}	\\[0.4ex]
\hline
A\(^2\Pi_{1/2}\)  	&	136.2	
	&	X\(^2\Sigma^+_{1/2}\) 	&	0	& 7.30 x10$^6$	&	99.461\%	\rule{0pt}{2.5ex}	\\
(\(J'' = 1/2\))
	&&	X\(^2\Sigma^+_{1/2}\) 	&	1	& 1.00 x10$^5$	&	 0.535\%	\rule{0pt}{2.5ex}	\\
	&&	X\(^2\Sigma^+_{1/2}\) 	&	2	& 1.26 x10$^1$	&	 0.001\%	\rule{0pt}{2.5ex}	\\
(\(J'' = 3/2\))
	&&	H\(^2\Delta_{3/2}\) 	&	0	& 4.24 x10$^2$	&	 0.003\%	\rule{0pt}{2.5ex}	\\[0.5ex]
\hline
B\(^2\Sigma^+_{1/2}\)	&	120.3
	&	X\(^2\Sigma^+_{1/2}\) 	&	0	& 8.21 x10$^6$	&	98.709\%	\rule{0pt}{2.5ex}	\\
 (\(J'' = 1/2\))
	&&	X\(^2\Sigma^+_{1/2}\) 	&	1	& 9.88 x10$^4$	&	 1.189\%	\rule{0pt}{2.5ex}	\\
	&&	X\(^2\Sigma^+_{1/2}\) 	&	2	& 3.83 x10$^2$	&	 0.005\%	\rule{0pt}{2.5ex}	\\
	&&	A\(^2\Pi_{1/2}\) 		&	0	& 1.74 x10$^3$	&	 0.021\%	\rule{0pt}{2.5ex}	\\
(\(J'' = 3/2\))
    &&	H\(^2\Delta_{3/2}\) 	&	0	& 4.44 x10$^3$	&	 0.053\%	\rule{0pt}{2.5ex}	\\	
	&&	A\(^2\Pi_{1/2}\) 		&	0	& 8.77 x10$^2$	&	 0.011\%	\rule{0pt}{2.5ex}	\\	
	&&	A\(^2\Pi_{3/2}\) 		&	0	& 1.04 x10$^3$	&	 0.012\%	\rule{0pt}{2.5ex}	\\[0.5ex]
\hline
H\(^2\Delta_{3/2}\)	&	5791
	&	X\(^2\Sigma^+_{1/2}\) 	&	0	& 1.02 x10$^5$  &	85.289\%	\rule{0pt}{2.5ex}	\\
 (\(N'' = 1\))
	&&	X\(^2\Sigma^+_{1/2}\) 	&	1	& 2.65 x10$^3$ 	&	 1.533\%	\rule{0pt}{2.5ex}	\\
	&&	X\(^2\Sigma^+_{1/2}\) 	&	2	& 1.70 x10$^1$	&	 0.010\%	\rule{0pt}{2.5ex}	\\
 (\(N'' = 3\))
	&&	X\(^2\Sigma^+_{1/2}\)	&	0	& 2.24 x10$^4$	&	 12.971\%	\rule{0pt}{2.5ex}	\\
	&&	X\(^2\Sigma^+_{1/2}\)	&	1	& 3.41 x10$^2$	&	  0.197\%	\rule{0pt}{2.5ex}	\\[0.6ex]		
\hline
\noalign{\vskip 2mm}
\end{tabular}}

\end{table}

The fraction of molecules that remain in the cooling cycle is determined by the number of loss channels that are optically linked to the excited state.
If all these decay channels are optically pumped then the cooling cycle is closed.
While common for atoms, this is unlikely in molecules because there are simply a larger number of decay pathways available.
The maximum number of cycles $N_n$ that $n$ light fields can support and maintain $f$ molecules in the cooling cycle is given by 

\begin{equation}
N_{n} \leq \frac{\log(f)}{\log(\sum_{k=1}^{n} ({\mathcal{R}_{ij}})_k)}
\end{equation}

\noindent where $\mathcal{R}_{ij}$ is the ratio of the Einstein A-coefficient for the vibronic transition $i - j$ to the total loss for the excited state \cite{Lane2015} and the sum is over those transitions that are optically pumped. The higher this sum (the closer to 1), the larger the number of cooling cycles that can be supported.
In this study decays to all possible electronic states are considered, not just the vibrational branching within a single electronic transition.
These additional decays are minute, typically less than 0.01\%, but become significant as the cooling cycle is extended.
Setting $f$ as 0.1 (90\% loss in molecular beam intensity) the number of cooling cycles supported by each cooling transition can be determined for both one and two colour cooling.
For one-colour cooling the number of cycles is small and very similar for the two transitions (Table~\ref{table3}), though A -- X is usually somewhat higher.
This is consistent with the largest decay channel being the 0-1 for both transitions. 

If this leak is plugged by a second laser, the B-state can still decay to the lower lying H$^2\Delta_{3/2}$, A$^2\Pi_{1/2}$ and A$^2\Pi_{3/2}$ states.
As a result the two-colour B -- X cooling strategy suffers approximately 0.102\% losses per cycle, limiting the number of cooling transitions to less than 2300 ($N_2$ = 2.25 x10$^3$).
By contrast, the two colour A -- X cooling transition supports over 53 thousand ($N_2$ = 5.37 x10$^4$) cycles.
Furthermore, multiple repumping transitions would be required to reactivate the cooling cycle using  B -- X while only one additional repump laser would be necessary to achieve the same using the alternative A$^2\Pi_{1/2}$ excited state.
The most important loss channel is to the  H\(^2\Delta\), \(v'' = \) 0 vibronic state (the level marked in Fig.~\ref{fig1}(b) by the lowest filled dot) for both the A$^2\Pi_{1/2}$  and, rather surprisingly, the B$^2\Sigma^+$ $v$ = 0 excited levels (indicated by the empty dots). Meanwhile, decay is also to both the A$^2\Pi_{1/2}$ and A$^2\Pi_{3/2}$ states when cooling using the B$^2\Sigma^+$ state.
The maximum deceleration, $a_{max}$ is slightly larger for the B -- X transition due to the shorter cooling wavelength and the slightly faster rate of decay but the superior cooling time ($N_2 \tau$) clearly ensures that pumping X$^2\Sigma^+_{1/2}$ ($v$ = 0, 1) into the A$^2\Pi_{1/2}$ state is the best cooling strategy.
Fig.~\ref{fig:Shift_effects} illustrates the sensitivity of both $N_1$ and $N_2$ (middle and right hand panels) to small changes in the upper state potential minimum and reveals that the $r_e$ dependence of these numbers can be very different, even within the same cooling transition.
With over fifty thousand cooling cycles in the two colour A -- X cooling transition, this technique could even cool the beam \cite{Iwata2017} of Iwata {\it et al.} down to the Doppler temperature using just two colours as around 3.7 x10$^4$ cycles are required at $a_{max}$.  

The population lost to the H$^2\Delta_{3/2}$ v = 0, $J$ = 3/2 level cannot exist indefinitely and ultimately decays radiatively to the ground X$^2\Sigma^+_{1/2}$ state.
As before, this transition is forbidden in Hund's case (a) but spin-orbit mixing results in some $^{2}\Pi$ character in the H$^2\Delta_{3/2}$ state.
The resulting intensity borrowing reduces the lifetime to $\tau$ = 5.8 $\mu$s.
This result is consistent with the experimental \cite{Bernard1987, Bernard1989, Barrow1991} observation of the weak H$^2\Delta_{3/2}$ $\rightarrow$ X$^2\Sigma^+_{1/2}$ transition.
This lifetime is much too long for strong, effective laser cooling, but its low Doppler temperature could suggest it might be very useful for producing and maintaining a very low temperature cloud of (already) trapped BaH radicals.
Unfortunately, there is no closed cycle for a $^2\Delta_{3/2}$ -- $^2\Sigma^+_{1/2}$ transition because the lowest excited rovibronic \(J' = 3/2\) state decays to two separate lower $N''$ levels (via the three available P-, Q- and R-branches).

\begin{table}
\centering
\caption{Calculated properties of the proposed laser cooling transitions in the BaH molecule.
For comparison the same parameters for the direct laser cooling of H atoms are displayed in the final column.
$N_i$ is the number of cycles supported by $i$ lasers before the population falls to 10\%.
$T_D$ and $v_D$ are the Doppler temperature and velocities, $T_r$ and $v_r$ are the recoil equivalents.
$v_c$ is the capture velocity.
The maximum acceleration (deceleration) is $a_{max} = \frac{\hbar k \gamma}{2M} = v_r \frac{A}{2}$.}
                            
\label{table3}
{\footnotesize
\begin{tabular}{|l|rrrc|}
\noalign{\vskip 5mm}
\hline
Molecular 		\rule{0pt}{2.5ex}	& \multicolumn{4}{|c|}{States}		\\
\hspace{3mm}data	&	B\(^2\Sigma^+_{1/2}\)	& A\(^2\Pi_{1/2}\)	&	H\(^2\Delta_{3/2}\) 	& H	\\
\hline
\(\lambda\)/nm$^a$	
	&	905.3 	&	1060.8	&	1110	&	121	
\rule{0pt}{2.5ex}	\\
\(\tau\)/ns 
	&	120.3 	&	136.2	&	5791	&	 1.6  
\rule{0pt}{2.5ex}	\\
\(N_1\)	
	&	177 	&	425	 &	-	&	$>$ $10^{18}$
\rule{0pt}{2.5ex}	\\
\(N_2\)	
	&	2.3~x$10^{3}$ 	&	5.4~x$10^{4}$	&	-	&$>$ $10^{20}$
\rule{0pt}{2.5ex}	\\
\(T_D\)/\(\mu\)K	
	&	31.7 	&	28.0	&	0.66	&	2349
\rule{0pt}{2.5ex}	\\
\(T_r\)/\(\mu\)K	
	&	0.168 	&	0.122	&	0.112	&	1285
\rule{0pt}{2.5ex}	\\
\(v_c\)/cm s\(^{-1}\)	
	&	119.9 	&	124.1	&	3.1	&	1211
\rule{0pt}{2.5ex}	\\
\(v_D\)/cm s\(^{-1}\)	
	&	4.36 	&	4.10	&	0.63	&	443
\rule{0pt}{2.5ex}	\\
\(v_r\)/cm s\(^{-1}\)	
	&	0.32 	&	0.27	&	0.26	&	325
\rule{0pt}{2.5ex}	\\
\(a_{max}\)/ms\(^{-2}\)	
	&	13166 	&	9932	&	-	&	1.0 x$10^9$
\rule{0pt}{2.5ex}	\\[1ex]
%	&	 	&		&		&	
%\rule{0pt}{0.1ex}	\\
\hline
\noalign{\vskip 1.8mm}
\multicolumn{3}{l}{\textsuperscript{a} Experimental wavelengths quoted.}\\
\end{tabular}}

\end{table}

\subsection{Further improvements to the cooling cycle}\label{Improv-cool}
The goal is clearly to ensure that the number of cooling cycles to achieve the Doppler temperature, $N_{Dopp}$, is much smaller than the number of cooling cycles that can be applied $N_{Dopp} \leq N_i$.
One method is to reduce $N_{Dopp}$ by reducing the initial velocity of the BaH beam.
Doyle and co-workers \cite{Lu2011} have demonstrated a buffer-gas cooled molecular beam of CaH with a forward velocity of just 65 ms$^{-1}$ so it may be possible to reduce the velocity further in the buffer-gas BaH beam.
Another approach is to use a Stark decelerator \cite{Meerakker2012} to reduce the beam velocity prior to laser cooling.
A travelling-wave design \cite{Osterwalder2010} is very effective at reducing the forward velocity without excessive beam losses.
The dipole moment of BaH has not been measured but an approximate value can be determined \cite{Bernath2015} using the method of Hou and Bernath that relies on using measured permanent dipole moments and equilibrium bond lengths of related ionic molecules.
For BaH, the relevant expression is based on these values for the ground state of the BaF radical

\begin{equation}
\mu_{D}(BaH)^2 R_{e}(BaH) = \mu_{D}(BaF)^2 R_{e}(BaF) - \phi
\end{equation}

\noindent where $\phi$ = 5.7202 D$^2$\AA, a constant based  \cite{Bernath2015} on the experimental properties of the CaF \cite{Huber1979} and CaH \cite{Steimle2008} radicals.
The estimated BaH dipole moment is 2.677 D, larger than both CaH and MgH though around 20\% smaller than corresponding fluoride.
Such a dipole moment is ideal for a travelling wave decelerator, particularly when combined with the low forward velocity of a BaH buffer-cooled molecular beam.
This estimated dipole is $\sim$20\% smaller than the permanent dipole at $r_e$, 3.32 D, computed here with the ACV$Q$Z basis set but less than a quarter the value calculated \cite{Lesiuk2017} by Lesiuk {\it et al}. 

An alternative approach is to add additional cooling lasers to plug the remaining leaks.
The most effective requires excitation out of the H\(^2\Delta_{3/2}\), $v$~=~0 level and repopulation of $v$~=~0 and $v$~=~1 in the X$^2\Sigma^+_{1/2}$ state.
The only suitable excited state appears to be 
E\(^2\Pi_{1/2}\) which requires laser radiation at $\approx$ 1775 nm (the vibronic levels involved are marked in Fig.~\ref{fig1}(b) by the filled dots).
At least 45 ro-vibronic levels can be populated by radiative decay from E$^2\Pi_{1/2}$ \(v = \) 0  (the transition dipole moments involving E\(^2\Pi_{1/2}\) as the upper state are shown in Fig.~ \ref{fig:cinf_TDM_all} (b)) so this repumping method would at first sight seem very inefficient.
However, unlike the situation with atomic levels, the FC factors help limit the number of strong transitions and this (in combination with the $\omega^3$ factor in the Einstein A-coefficient for a diatomic transition \cite{Hansson2005}) ensures that over 94\% (94.61\%) of the decay is back into the cooling cycle (the details can be found in Table~\ref{tab:E1_decays}).
The calculated lifetime of the E$^2\Pi_{1/2}$ \(v = \) 0 state is 45.5 ns.
This effectively removes the H\(^2\Delta_{3/2}\) state as the main loss channel (which becomes decay to X$^2\Sigma^+_{1/2}$  \(v'' = \) 2 instead) and increases the number of cooling cycles towards a quarter of a million ($N_{3}$ = 2.34 x10$^5$).

\begin{table*}
\small
\centering
\caption{Decay routes from E$^2\Pi$ \(_{1/2}\) (\(v\)=0, \(J = \frac{1}{2}\)) with energies determined from Duo data alone ($\omega_{Duo}$) and with corrections using experimental data ($\omega_{exp}$). $\Delta$ is the \% difference between $\omega_{Duo}$ and $\omega_{exp}$.
Decay branching ratios using the corrected energies are also tabulated (only decays of 0.01\% or greater shown).
Despite at least 45 decay paths available, nearly 94\% returns to X\(^2\Sigma^+_{1/2}\) \(v\) = 0 and 1. $\mathcal{A}$ is the Einstein A-coefficient for each transition and $\mathcal{R}$atio is the value of $\mathcal{R}_{ij}$ in percentage terms.}
\label{tab:E1_decays}
\begin{tabular}{|ll|rrr|rr|}
\noalign{\vskip 4mm}
\hline
Final State	&	\(v\) \rule{0pt}{2.5ex}
	& \(\omega_{Duo}\)\hspace{2mm}  
	& \(\omega_{exp}\)\hspace{2.4mm} & $\Delta$ \hspace{2mm} & \(\mathcal{A}\) / s\(^{-1}\)\hspace{1.4mm} & $\mathcal{R}$atio\hspace{1.8mm}	\\[.8ex] \hline
X\(^2\Sigma^+_{1/2}\), \(N=1\) \rule{0pt}{2.5ex}
  & 0 &  14695.52 & 14625.57 & 0.5\% & 1.99 x10$^7$ & 91.01\% \\
  & 1 &  13565.12 & 13486.41 & 0.6\% & 6.49 x10$^5$ &  2.97\% \\
  & 2 &  12462.62 & 12376.21 & 0.7\% & 2.57 x10$^4$ &  0.12\% \\
  & 3 &  11387.93 & 11294.84 & 0.8\% & 1.73 x10$^3$ &  0.01\% \\[.8ex] \hline
H\(^2\Delta_{3/2}\), \(J=\frac{3}{2}\) \rule{0pt}{2.5ex}
  &  0  &  5273.09  &  5648.33  &  -6.6\%  & 1.38 x10$^5$  &  0.63\%  \\
  &  1  &  4216.00  &  4583.95  &  -8.0\%  & 8.10 x10$^3$  &  0.04\%  \\[.8ex] \hline
A\(^2\Pi_{1/2}\), \(J=\frac{1}{2}\) \rule{0pt}{2.5ex}
  &  0  &  4934.19  &  5198.72  &  -5.1\%  & 1.43 x10$^5$  &  0.65\%  \\
  &  1  &  3856.74  &  4118.89  &  -6.4\%  & 8.87 x10$^3$  &  0.04\%  \\[.8ex]
A\(^2\Pi_{1/2}\), \(J=\frac{3}{2}\) \rule{0pt}{2.5ex}
  &  0  &  4915.83  &  5186.40  &  -5.2\%  & 8.19 x10$^5$  &  3.75\%  \\
  &  1  &  3838.77  &  4106.78  &  -6.5\%  & 1.68 x10$^4$  &  0.08\%  \\
  &  2  &  2791.66  &  3057.76  &  -8.7\%  & 1.40 x10$^3$  &  0.01\%  \\[.8ex]
A\(^2\Pi_{3/2}\), \(J=\frac{3}{2}\) \rule{0pt}{2.5ex}
  &  0  &  4427.07  &  4709.22  &  -6.0\%  & 3.31 x10$^3$ &  0.02\% \\[.8ex] \hline
B\(^2\Sigma^+_{1/2}\), \(J=\frac{1}{2}\) \rule{0pt}{2.5ex}
  &  0  &  3522.47  &  3568.39  &  -1.3\%  & 1.21 x10$^5$ &  0.56\%  \\
     \qquad\quad(\(N=1\))
  &  1  &  2462.33  &  2510.62  &  -1.9\%  & 4.83 x10$^3$ &  0.02\%  \\[.8ex]
B\(^2\Sigma^+_{1/2}\), \(J=\frac{3}{2}\) \rule{0pt}{2.5ex}
  &  0  &  3532.54  &  3575.52  &  -1.2\%  & 1.97 x10$^4$ &  0.09\%  \\
   \qquad\quad(\(N=1\))
  &  1  &  2472.25  &  2517.57  &  -1.8\%  & 1.26 x10$^3$ &  0.01\%  \\[.8ex] \hline
\end{tabular}
\end{table*}

The above analysis may still be somewhat of an idealisation because it assumes (1) there are only electric dipole decay pathways and (2) that the primary two cooling lasers use the same upper level. When the decay level lies around 0.01\% there is a possibility that magnetic dipole \cite{Kirste2012} and electric quadrupole transitions can take place that preserve the parity and therefore break the cooling cycle. Meanwhile, the use of two different excited vibronic levels helps prevent any possible interference effects suppressing absorption but inevitably brings the problem of additional decay pathways from the new excited level.
The worst case here would be to adopt the strong B -- X (1 - 1) diagonal transition instead of A -- X (0 - 1) considered in Section \ref{Compare-cool} because this not only enhances the  decay to X$^2\Sigma^+_{1/2}$ \(v'' = \) 2 (this now exceeds the loss to the H$^2\Delta_{3/2}$ state) but also introduces losses  to \(v'' = \) 3.
This effectively increases the losses further by

\begin{equation}
\approx {\mathcal{R}^{A-X}_{01}}[1 - ({\mathcal{R}^{B-X}_{10}} + {\mathcal{R}^{B-X}_{11}})]
\end{equation}

The total repumping losses now lie above 2.6\% (as opposed to essentially zero in the earlier model, see Table~\ref{table4}) resulting in the overall two-colour loss rising to almost 0.015\%, ten times that in the ideal A -- X laser cooling scenario.
Note how the lifetime of this level is slightly longer than \(v' = \) 0.
The lowest losses are achieved using the B -- X (0 -- 1) branch transition instead as this limits the increased decay to around 0.0005\%.
However, even this small increase reduced the value of $N_2$ by over six thousand cooling cycles ($N_{2}$ = 4.77 x10$^4$) and the three-colour cycles to $N_{3}$ = 1.51 x10$^5$, down by almost one hundred thousand.
This suggests that an additional improvement would be to keep the shared upper rovibronic level for the two cooling lasers but modulate the laser amplitudes in anti-phase in order to prevent dark state formation.

\begin{table}
\centering
\caption{Radiative decay pathways from the B$^2\Sigma^+$ $v^\prime$ = 1 level of the BaH molecule. \(v''\) is the final vibrational level following decay. The calculated radiative lifetime $\tau$ is 125.2 ns.}
                                           
\label{table4}
{\footnotesize
\begin{tabular}{|l|llrr|}
\noalign{\vskip 3mm}
\hline 
	& \multicolumn{4}{|c|}{Decay pathways}		\\
    & Final State  & \(v''\)	& \(\mathcal{A}\) / s\(^{-1}\)\hspace{1mm}	& $\mathcal{R}$atio\hspace{1.7mm}	\\[0.2ex]
\hline
B\(^2\Sigma^+_{1/2}\)		
	&	X\(^2\Sigma^+_{1/2}\) 	&	0	&  3.32 x10$^5$	&	 4.16\%	\rule{0pt}{2.5ex}	\\
\(v' =\) 1,
	 &	X\(^2\Sigma^+_{1/2}\) 	&	1	&  7.45 x10$^6$	&	93.29\%	\rule{0pt}{2.5ex}	\\
 \(J' = 1/2\)
	&	X\(^2\Sigma^+_{1/2}\) 	&	2	&  1.94 x10$^5$	&	 2.43\%	\rule{0pt}{2.5ex}	\\
	&	X\(^2\Sigma^+_{1/2}\) 	&	3	&  1.14 x10$^3$	&	 0.01\%	\rule{0pt}{2.5ex}	\\
	&	H\(^2\Delta_{3/2}\) 	 & 1	&  5.08 x10$^3$	&	 0.06\%	\rule{0pt}{2.5ex}	\\
	&	A\(^2\Pi_{1/2}\) 		&	1	&  2.40 x10$^3$	&	 0.03\%	\rule{0pt}{2.5ex}	\\
	&	A\(^2\Pi_{3/2}\) 		&	1	&  9.54 x10$^2$	&	 0.01\%	\rule{0pt}{2.5ex}	\\[0.6ex]
	
\hline
\noalign{\vskip 2mm}
\end{tabular}}

\end{table}

%----------------------------------------------------------------
\section{
\label{sec:concl}
Conclusions}
%----------------------------------------------------------------

In many ways the simulation of laser cooling dynamics is one of the most stringent tests of {\it ab initio} quantum chemistry by virtue of the thousands of transitions that must be successfully computed.
This paper has highlighted a number of these issues in the case of the radical hydride BaH.
By including spin-orbit coupling to the analysis of laser cooling at the ro-vibrational level, it is clear that the redder A\(^2\Pi\) -- X\(^2\Sigma^+ \) cooling transition is preferable to the alternative B\(^2\Sigma^+ \) -- X\(^2\Sigma^+ \) despite the longer excited state lifetime (136 vs.
120~ns).
A further new feature is the appearance of losses at the 0.05\% level via B\(^2\Sigma^+_{1/2}\) \(\rightarrow\) H\(^2\Delta_{3/2}\) spontaneous decay.
It should prove possible to cool a buffer-gas cooled beam of BaH down to the Doppler temperature with just two cooling lasers.
However, quantitative information (such as the maximum number of cooling cycles) is difficult to extract from the {\it ab initio} calculations, even with the help of crucial experimental data. 

%----------------------------------------------------------------
\section{
\label{sec:ackn}
Acknowledgments}
%----------------------------------------------------------------

We thank Romain Garnier for help with the initial calculations and Tanya Zelevinsky for useful discussions on the experimental laser cooling of BaH.
We express our gratitude for the financial support of the Leverhulme Trust (Research Grant RPG-2014-212) including the funding of a studentship for KM.

%\clearpage
\bibliographystyle{unsrt}
%\bibliography{BaHDecays_refs}

\end{thebibliography}

\end{document}